\documentclass[%
reprint,
amsmath,amssymb,
aps,superscriptaddress,
prapplied,longbibliography,
floatfix,
]{revtex4-2}

\usepackage{graphicx}% Include figure files
\usepackage{dcolumn}% Align table columns on decimal point
\usepackage{bm}% bold math
\usepackage{hyperref}

\hypersetup{
	colorlinks   = true, %Colours links instead of ugly boxes
	urlcolor     = blue, %Colour for external hyperlinks
	linkcolor    = blue, %Colour of internal links
	citecolor    = blue %Colour of citations
}

\usepackage{soul}

\begin{document}
	
\preprint{APS/123-QED}
	
\title{Limits to Quantum %two-spin-qubit 
Gate Fidelity from Near-Field Thermal and Vacuum Fluctuations}% Force line breaks with \\

\author{Wenbo Sun}
 \affiliation{Elmore Family School of Electrical and Computer Engineering, Purdue University, West Lafayette, Indiana 47907, USA}
 \affiliation{Birck Nanotechnology Center, Purdue University, West Lafayette, Indiana 47907, USA}
\author{Sathwik Bharadwaj}
 \affiliation{Elmore Family School of Electrical and Computer Engineering, Purdue University, West Lafayette, Indiana 47907, USA}
 \affiliation{Birck Nanotechnology Center, Purdue University, West Lafayette, Indiana 47907, USA}
 \author{Li-Ping Yang}
 \affiliation{Center for Quantum Sciences and School of Physics, Northeast Normal University, Changchun, China}
 \author{Yu-Ling Hsueh}
 \affiliation{Silicon Quantum Computing Pty Ltd., University of New South Wales, Sydney, NSW, 2052, Australia}
 \affiliation{School of Physics, The University of New South Wales, Sydney, NSW 2052, Australia}
 \author{Yifan Wang}
 \affiliation{Elmore Family School of Electrical and Computer Engineering, Purdue University, West Lafayette, Indiana 47907, USA}
 \author{Dan Jiao}
 \affiliation{Elmore Family School of Electrical and Computer Engineering, Purdue University, West Lafayette, Indiana 47907, USA}
 \author{Rajib Rahman}
 \affiliation{Silicon Quantum Computing Pty Ltd., University of New South Wales, Sydney, NSW, 2052, Australia}
 \affiliation{School of Physics, The University of New South Wales, Sydney, NSW 2052, Australia}
 \author{Zubin Jacob}
 \email{zjacob@purdue.edu}
 \affiliation{Elmore Family School of Electrical and Computer Engineering, Purdue University, West Lafayette, Indiana 47907, USA}
 \affiliation{Birck Nanotechnology Center, Purdue University, West Lafayette, Indiana 47907, USA}

	\date{\today}% It is always \today, today,
	%  but any date may be explicitly specified
	
	\begin{abstract}
		High-fidelity quantum gate operations are essential for achieving scalable quantum circuits. In spin qubit quantum computing systems, metallic gates and antennas that are necessary for qubit operation, initialization, and readout, also cause detrimental effects by enhancing fluctuations of electromagnetic fields. Therefore, evanescent wave Johnson noise (EWJN) caused by near-field thermal and vacuum fluctuations becomes an important unmitigated noise, which induces the decoherence of spin qubits and limits the quantum gate operation fidelity. Here, we first develop a macroscopic quantum electrodynamics theory of EWJN to account for the dynamics of two spin qubits interacting with metallic circuitry. Then we propose a numerical technique based on volume integral equations to quantify EWJN strength in the vicinity of nanofabricated metallic gates with arbitrary geometry. We study the limits to two-spin-qubit gate fidelity from EWJN-induced relaxation processes in two experimentally relevant quantum computing platforms: (a) the silicon quantum dot system and (b) nitrogen-vacancy centers in diamond. Finally, we introduce a Lindbladian engineering method to optimize the control pulse sequence design and show its enhanced performance over Hamiltonian engineering in mitigating the influence of thermal and vacuum fluctuations. Our work leverages advances in computational electromagnetics, fluctuational electrodynamics, and open quantum systems to suppress the effects of near-field thermal and vacuum fluctuations and reach the limits of two-spin-qubit gate fidelity. 
	\end{abstract}
	
	%\keywords{Suggested keywords}%Use showkeys class option if keyword
	%display desired
	\maketitle
	
	%\tableofcontents
	
	\section{Introduction}\label{section1}
	
	Quantum gate fidelity measures closeness between physically implemented quantum gate operations and theoretically ideal counterparts. Realizing high-fidelity quantum gate operations is necessary for any practical application of quantum computing. Recent demonstrations of quantum gate operations have observed increasing fidelity by mitigating various noise sources and shown spin qubits in multiple solid-state systems as promising candidates for future quantum computers \cite{Xue2022threshold,Noiri2022,Hendrickx2021,PhysRevLett.124.220501,Adam2022two,Huang2019fbenchmark,Wu2019,Rong2015}. The current record for two-spin-qubit gate fidelity has exceeded $99.5\%$ \cite{Xue2022threshold,Noiri2022,Adam2022two,Hendrickx2021,Rong2015}.
	
	Despite these significant achievements, pushing the current state-of-the-art quantum gate fidelity to higher records is still central for quantum computing research. The complexity of scalable quantum circuits is sensitive to the fidelity of underlying two-qubit quantum gate operations. This arises since lower gate fidelity causes the need for a  larger number of qubits in the error correction process. Error correction is critical for eliminating uncertainty in the final results of quantum circuits. For the case of one popular error correction method - surface code \cite{Campbell2017,PhysRevA.86.032324,PhysRevA.83.020302}, increasing gate fidelity from $99.9\%$ to $99.999\%$ could decrease the number of qubits required for building the same quantum circuits by a factor of $10$ \cite{PhysRevA.86.032324}.
	
\begin{figure}
	    \centering
	    \includegraphics[width = 3.2in]{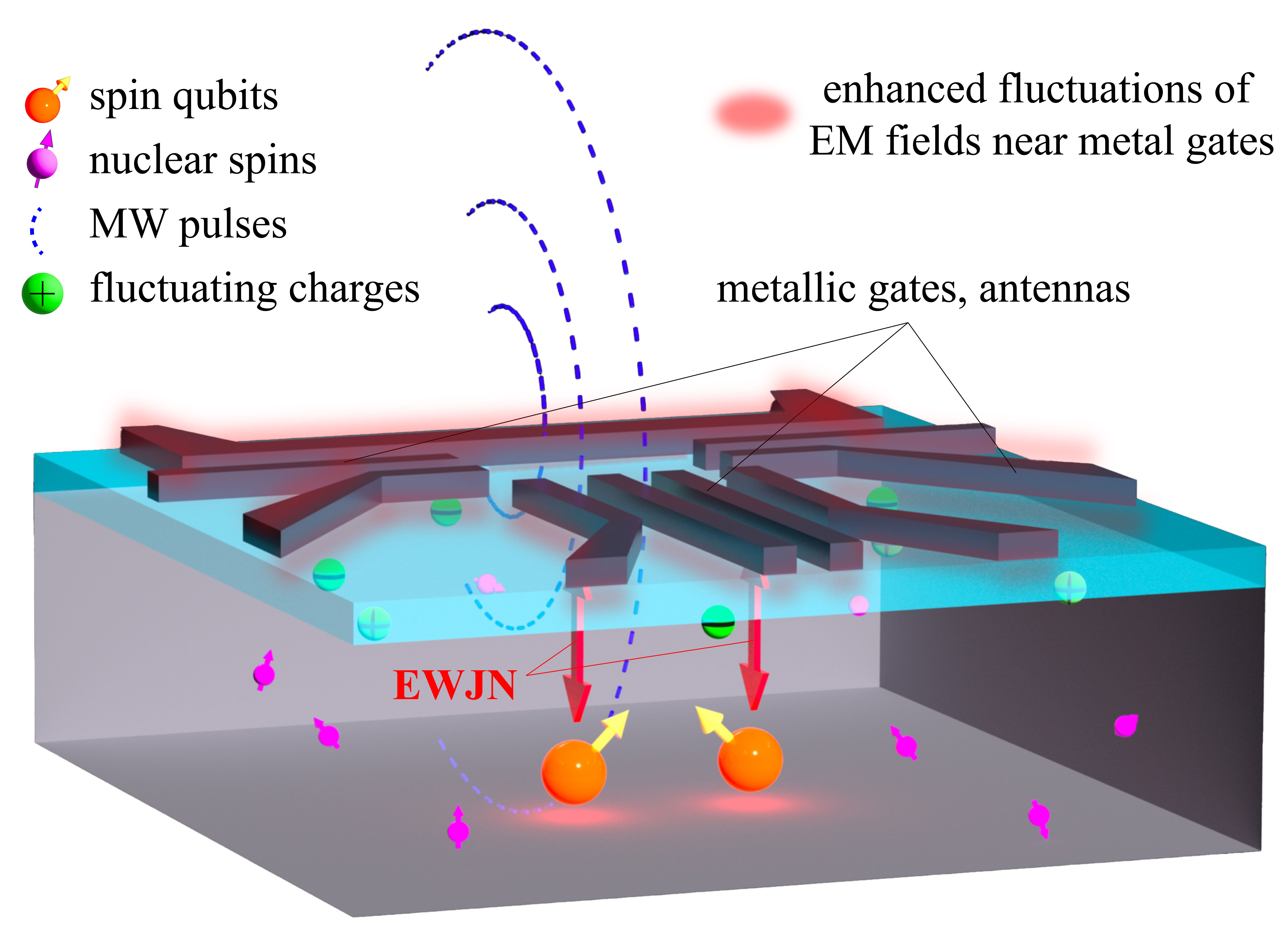}
	    \caption{Schematic of a spin qubit quantum computing system with different noise sources. Four kinds of noise are shown: nuclear spin bath noise, noise in control pulses, charge noise, and evanescent wave Johnson noise (EWJN). Red arrows represent interactions between spin qubits and enhanced vacuum and thermal fluctuations of electromagnetic fields.}
	    \label{fig:fig1}
	\end{figure}

	Environmental noise from various origins is the roadblock to achieving high gate fidelity. Figure~\ref{fig:fig1} shows the schematic of a spin qubit quantum computing device with different types of noise. Among all the noise sources, near-field thermal and vacuum fluctuations are important but are relatively less explored. Previous works have quantified the influence of various other noise sources and demonstrated corresponding optimization methods (table \ref{tab:table1}). For instance, nuclear spin noise originating from non-zero nuclear spins of silicon isotopes $\mathrm{^{29}Si}$ can decohere spin qubits \cite{Chekhovich2013,PhysRevB.74.035322}. However, with recent improvements in materials and fabrication technology, nuclear spin noise is suppressed in isotopically enriched $\mathrm{^{28}Si}$ samples \cite{itoh_watanabe_2014}. Another important fluctuation is induced by instability in microwave pulses used for spin qubits control. One potential solution is redesigning the amplitudes and lengths of microwave pulses \cite{Rong2015}. Fluctuating charges in semiconductor quantum dots can lead to charge noise that induces spin qubits' decay~\cite{kuhlmann2013charge,PhysRevB.100.165305}. Biasing spin qubits symmetrically reduces their sensitivity to charge noise, thus decreasing the impacts of charge noise in quantum gate operations \cite{PhysRevLett.116.110402,PhysRevLett.116.116801,Abadillo-Uriel2019}. In contrast, the influence of near-field thermal and vacuum fluctuations of electromagnetic (EM) fields on quantum gate fidelity is not well evaluated or mitigated.
	
	In the quantum computing system based on spin qubits, metallic lines, gates, and antennas are necessary for initialization, control, and readout \cite{Hendrickx2021,Noiri2022,Huang2019fbenchmark,PhysRevLett.124.220501}. Error correction protocols rely on metallic lines for large-scale qubit manipulations as well \cite{Li2018siliconnetwork,Chen2020diamondprocessor}. Evanescent surface and bulk waves in these metals inside the quantum computing devices intensify the fluctuations of EM fields \cite{FORD1984195}. Therefore, evanescent wave Johnson noise (EWJN) originating from near-field thermal and vacuum fluctuations in the vicinity of metals is inevitable in quantum computing devices. 
	
	Multiple experimental and theoretical results show EWJN as an important and prevalent type of noise that induces spin qubit relaxation processes in various devices.  In the semiconductor quantum dot system, theoretical studies have shown signatures of EWJN effects on the relaxation time (T1) \cite{Langsjoen2012ewjn,PhysRevB.87.045301,PhysRevA.103.062401}. Experiments on silicon quantum dot qubits display a deviation of the measured spin relaxation rate from the expected phonon-induced relaxation rate at low magnetic fields \cite{Yang2013}, which can be explained by Johnson noise \cite{PhysRevB.90.235315}. A recent experiment on donor qubits in metal oxide semiconductor (MOS) devices also shows that the measured spin relaxation rate deviates from the typical $B^5$ dependency, a signature of spin-orbit mediated phonon relaxation, and follows a linear $B$ dependency at low magnetic fields ($B<3~\mathrm{T}$) \cite{Tenberg2019siliconrelaxation}. Meanwhile, EWJN has been proposed to give this linear B dependency \cite{Tenberg2019siliconrelaxation}. In the diamond NV center system, the measured relaxation time of NV electron spin qubits in the vicinity of metal films matches well with the Ford-Weber EWJN theory \cite{kolkowitz2015johnsonnoise,FORD1984195}. The Ford-Weber theory of understanding EWJN and the non-local dielectric response of metals can also be used to derive bounds for the qubit relaxation rate near thin films. However, current studies fail to provide accurate EWJN-induced decay rates in the vicinity of metallic gates with realistic geometry. In addition, a thorough study of EWJN-induced cooperative effects in a near-resonant multi-spin-qubit system is elusive, although two-point noise correlation functions near metallic objects with simple geometries have been evaluated~\cite{Premakumar_2018, PhysRevA.103.062401}. Furthermore, quantum dynamics of the qubit system including EWJN effects cannot be captured by conventional Ford-Weber theory. Thus moving beyond Ford-Weber EWJN theory is necessary for understanding two-qubit quantum dynamics and optimizing realistic gate geometries. 
	
    \begin{table}
	\caption{\label{tab:table1}
	Different noise sources and corresponding optimization methods.}
	\begin{ruledtabular}
    \begin{tabular}{cccccccc}
     noise source & optimization method \\
    \hline\\[-0.8em]
    nuclear spin bath & isotopic enrichment \cite{itoh_watanabe_2014} \\[+0.2em]
    imperfect MW pulses & Hamiltonian engineering \cite{Rong2015} \\[+0.2em]
    fluctuating charges & symmetric operation \cite{PhysRevLett.116.110402,PhysRevLett.116.116801,Abadillo-Uriel2019} \\[+0.2em]
    \hline\\[-0.8em]
    \begin{tabular}{@{}c@{}} thermal and vacuum \\ fluctuations \end{tabular} & \begin{tabular}{@{}c@{}} Lindbladian engineering \\ (this work) \footnote{Lindbladian engineering is also applicable to other Markovian noise} 
    \end{tabular} \\
    \end{tabular}
    \end{ruledtabular}
    \end{table}
    
	Near-field thermal and vacuum fluctuations require a careful treatment distinct from aspects like nuclear spin noise to mitigate their influence on quantum gate fidelity. Metallic systems that enhance thermal and vacuum fluctuations are essential and nearly irreplaceable for quantum gate operations. Lowering operation temperature reduces the influence of thermal fluctuations; however, vacuum fluctuations persist even at $T = 0~\mathrm{K}$. A feasible solution is optimizing microwave pulse sequences used for qubit control. Hamiltonian engineering and dynamical decoupling are popular techniques to seek optimal control of quantum systems \cite{OKeeffe2019hamiltonian,choi2020hamiltonian,Yang2011,PhysRevLett.105.230503, Du2009, doi:10.1126/science.1192739, PhysRevLett.104.130501}. In Hamiltonian engineering, unwanted interactions represented by Hermitian Hamiltonian operators are effectively suppressed in rotating frames corresponding to microwave pulses. However, thermal and vacuum fluctuations and other Markovian noise are usually captured in a non-Hermitian super-operator \cite{breuer2002theory,Schulte_Herbr_ggen_2011}. Apart from that, implementations of Hamiltonian engineering and dynamical decoupling commonly focus on reducing noise from the spin bath instead of near-field thermal and vacuum flutuations. 
	
	In this paper, we first provide a macroscopic quantum electrodynamics (QED) theory of EWJN. Based on macroscopic QED \cite{scheel2009macroqed,cris2022fbound,Yang_2020}, we obtain the Lindblad master equation, which describes the non-unitary dynamics of the two-spin-qubit system coupled to vacuum and thermal fluctuations. We consider contributions from all spontaneous, stimulated, and cooperative decay processes induced by EWJN. The cooperative decay appears naturally in the multi-qubit system. In the quantum computing system based on spin qubits, cooperative decay could also happen at a comparable rate as stimulated and spontaneous decay, indicating it as an important part of EWJN that has not been well studied in previous works. Similar methods have been used to study quantum cooperative effects near metamaterials and single photon pulse induced transient entanglement near graphene \cite{cris2022fbound,Yang_2020}.
	
	Previous studies have only considered the thin film geometry of metal gates and its effect on spin qubits. Here, we overcome this stumbling block and consider realistic gate geometry by combining computational electromagnetics approaches with the above-mentioned quantum theory. We introduce the volume integral equations (VIEs) method as an efficient and accurate numerical technique to calculate spontaneous and cooperative decay rates of spin qubits in the vicinity of metallic gates with arbitrary geometry. The VIE-based method has great flexibility in modeling complicated geometry in open-region settings. Non-local effects also have important contributions to decay rates when the operating temperature is low, and the distance between qubits and metallic gates is small \cite{kolkowitz2015johnsonnoise,Langsjoen2012ewjn}. As a result, we take non-local effects into calculations when applicable. 
	
	Next, we evaluate limits to controlled-NOT (CNOT) gate fidelity due to vacuum and thermal fluctuations induced spin qubit relaxation processes in two popular quantum computing systems: NV center in diamond and quantum dot in silicon. CNOT gate is a representative two-qubit quantum gate involving interaction between two qubits. Its fidelity is commonly used to assess the performance of quantum computing devices \cite{Veldhorst2015,PhysRevLett.124.220501}. We calculate the CNOT gate fidelity through system dynamics simulations based on the Lindblad master equation. 
	
	Finally, we propose a Lindbladian engineering method as an effective approach to reduce the influence of near-field thermal and vacuum fluctuations. Our Lindbladian engineering approach searches for optimal control pulse sequences via dynamics determined by the Lindblad master equation. We provide optimized control protocols robust against Markovian relaxation and dephasing processes through Lindbladian engineering targeting two-spin-qubit systems.
	
	The paper's outline is as follows. In section~\ref{section2}, we introduce the macroscopic QED theory of EWJN and important numerical techniques. Then, in section~\ref{section3} and section~\ref{section4}, we evaluate EWJN's influence on CNOT gate fidelity in the silicon quantum dot and NV center system. Next, in section~\ref{section5}, we present the application of Lindbladian engineering in both systems with Markovian relaxation and dephasing processes of spin qubits. Finally, in section~\ref{section6}, we indicate future prospects for further research.
	
	\section{Macroscopic Quantum electrodynamics theory of thermal and vacuum fluctuations and Lindbladian Engineering}\label{section2}

	In this section, we first discuss the dynamics of the two-spin-qubit system and the Lindblad master equation. We adopt macroscopic quantum electrodynamics (QED) approaches \cite{scheel2009macroqed,scheel2009arxivqed} to study EWJN effects in the vicinity of metal gates. %in system dynamics. 
	We present the analytical and numerical methods to obtain spontaneous and cooperative decay rates of spin qubits induced by thermal and vacuum fluctuations. We elucidate how geometries and non-local dielectric responses of the metal gates, and cooperative effects influence the decoherence of the two-spin-qubit system. Finally, we demonstrate the Lindbladian engineering for control pulse optimization to mitigate the influence of EWJN.
	
	\subsection{Macroscopic QED theory of EWJN}
	
	In our model, the system consists of two spin qubits coupled to each other via the dipole-dipole or exchange interaction. Two-qubit quantum gate operations are realized by driving the qubits using resonant microwave pulses. Near-field vacuum and thermal fluctuations of EM fields induce spontaneous, stimulated, and cooperative decay processes through the interaction between the system and the Markovian photon bath environment. Macroscopic QED theory valid in arbitrary dissipative media \cite{scheel2009macroqed,scheel2009arxivqed} provides a more accurate description of the system-bath interaction and system dynamics compared to semi-classical methods such as Fermi's golden rule and Ford-Weber theory, which fail to adequately describe cooperative decay processes \cite{svidzinsky2008Fermis,rotter2015review}. In the Schrodinger picture, following the quantization framework in macroscopic QED~\cite{scheel2009macroqed,Yang_2020,cris2022fbound}, the total Hamiltonian describing the two-spin-qubit system and the electromagnetic bath can be written as:
    \begin{multline}\label{totalH}
        \hat{H}_{tot}=\sum_{i=1}^2 \hbar \omega_{i} \ \hat{\sigma}_i^+ \hat{\sigma}_i^- + \int d^3 \mathbf{r} \int_0^\infty d \omega \, \hbar \omega \, \hat{\mathbf{f}}^\dagger (\mathbf{r},\omega) \hat{\mathbf{f}}(\mathbf{r},\omega) \\
        - \sum_{i=1}^2 (\mathbf{m}_{i}\hat{\sigma}_i^+ + \mathbf{m}_{i}^\dagger\hat{\sigma}_i^-) \cdot \hat{\mathbf{B}}(\mathbf{r}_i)+ \hat{H}_{ex}.
    \end{multline} 

    Here, $\omega_i, \mathbf{m}_i, \mathbf{r}_i$ stand for the resonance frequency, spin magnetic dipole moment, and position of the $i \mathrm{th}\ (i=1,2)$ spin qubit. $\hat{\sigma}_i^{+(-)}$ represents the raising (lowering) operator with respect to the $i \mathrm{th}$ spin qubit. $\hat{\mathbf{f}}^\dagger(\mathbf{r},\omega)$ and $\hat{\mathbf{f}}(\mathbf{r},\omega)$ denote the photon/polariton creation and annihilation operators. $\hat{\mathbf{B}}(\mathbf{r}_i)$ is the magnetic field operator at the qubit position and can be written in terms of the dyadic Green's functions $\overleftrightarrow{G}$ and $\hat{\mathbf{f}}^\dagger, \hat{\mathbf{f}}$~\cite{cris2022fbound,scheel2009arxivqed,Yang_2020} (Appendix.~\ref{qedtheory}). On the right-hand side of Eq.~(\ref{totalH}), the first and second terms represent the Hamiltonians of the two qubits and the electromagnetic bath separately. The third term describes the interaction between the spin qubits and the fluctuating electromagnetic bath. Here, we adopt the dipole approximation widely used for studying EWJN and the interaction of fluctuating electromagnetic fields with solid-state qubits~\cite{kolkowitz2015johnsonnoise,Tenberg2019siliconrelaxation,PhysRevLett.110.080502,PhysRevLett.94.227402,Lodahl2004}. High-order corrections from the large spatial extension of qubits and the relativistic effects are neglected in this manuscript but can be studied similarly by modifying the interaction Hamiltonian~\cite{PhysRevA.37.2284,PhysRevB.86.085304}. The last term $\hat{H}_{ex}$ stands for the exchange coupling Hamiltonian between the exchange-coupled spin qubits originating from the Coulomb interaction and the Pauli exclusion principle~\cite{PhysRevB.59.2070}. When the two spin qubits are not coupled by the exchange interaction, the exchange coupling strength $J$ in $\hat{H}_{ex}$ is considered to be 0.
    
    With the Born-Markovian approximation, we can derive the Lindblad master equation governing the time evolution of the two-spin-qubit density matrix $\rho_q(t)$ from Eq.~(\ref{totalH}) (see Appendix.~\ref{qedtheory} for derivations):
	
	\begin{align}\label{mainlindbladT}
	\begin{aligned}
	    &\frac{d\rho_q(t)}{dt}=\frac{1}{i \hbar} [\hat{H}(t),\rho_{q}(t)] + \hat{\hat{L}}_r \rho_{q}(t) = \frac{1}{i \hbar} [\hat{H}(t),\rho_{q}(t)]  \\
        &+ \sum_{i,j}  \Gamma_{ij}  [\hat{\sigma}_i^- \rho_q(t) \hat{\sigma}_j^+ - \frac{1}{2} \rho_q(t) \hat{\sigma}_i^+ \hat{\sigma}_j^- - \frac{1}{2} \hat{\sigma}_i^+ \hat{\sigma}_j^- \rho_q(t)] \\
        &+ \sum_{i,j} \mathcal{N}_{ij}  \Gamma_{ij}  [\hat{\sigma}_i^- \rho_q(t) \hat{\sigma}_j^+ - \frac{1}{2} \rho_q(t) \hat{\sigma}_i^+ \hat{\sigma}_j^- - \frac{1}{2} \hat{\sigma}_i^+ \hat{\sigma}_j^- \rho_q(t)] \\
        & + \sum_{i,j} \mathcal{N}_{ij} \Gamma_{ij}  [\hat{\sigma}_i^+ \rho_q(t) \hat{\sigma}_j^- 
        - \frac{1}{2} \rho_q(t) \hat{\sigma}_i^- \hat{\sigma}_j^+ - \frac{1}{2} \hat{\sigma}_i^- \hat{\sigma}_j^+ \rho_q(t)],
	\end{aligned}
    \end{align}
    where $i,j \in \{1,2\}$. $\hat{H}(t)$ is the Hamiltonian governing the unitary evolution of $\rho_q(t)$. It consists of control Hamiltonians $\hat{H}_{mw}$ corresponding to microwave pulses and the coupling Hamiltonian between spin qubits. Coupling Hamiltonians of spin qubits can be dipole-dipole interaction $\hat{H}_{dd}$ dominant in the NV center system, or exchange coupling Hamiltonian $\hat{H}_{ex}$ dominant in the silicon quantum dot system. $\hat{\hat{L}}_r$ is the trace-preserving Lindblad super-operator. It describes the non-unitary relaxation processes within the computational subspace that are induced by near-field thermal and vacuum fluctuations. $\hat{\sigma}_i^{+(-)}$ is the raising (lowering) operator for the $i \mathrm{th}$ spin qubit. $\Gamma_{ij}$ represents the spontaneous ($i=j$) and cooperative ($i\neq j$) decay rates, which will be discussed in detail in the next subsection. $\mathcal{N}_{ij}$ is the mean photon number at thermal bath temperature $T$ and average spin qubit resonance frequency $\omega_{+}=\frac{\omega_i + \omega_j}{2}$:
	
	\begin{equation}\label{meanphotonnumber}
		\mathcal{N}(\omega_+,T)=\frac{1}{e^{\frac{\hbar \omega_+}{k_BT}}-1}.
	\end{equation}

	Here, we briefly discuss Eq.~(\ref{mainlindbladT}) and different terms in $\hat{\hat{L}}_r$. Eq.~(\ref{mainlindbladT}) is valid for two qubits with resonance frequencies $|\omega_i - \omega_j| \ll \omega_i + \omega_j$. We have taken a Born-Markovian approximation to obtain Eq.~(\ref{mainlindbladT}), where we assume that the two-spin-qubit system is weakly coupled to the thermal photon bath (Born approximation) and the bath correlation time $\tau_c$ is much smaller than the relaxation times of the system $\tau_c \ll \Gamma_{ij}^{-1}$ (Markovian approximation) \cite{breuer2002theory} (see Appendix.~\ref{qedtheory} for more details). The second term on the right-hand side of Eq.~(\ref{mainlindbladT}) describes the spontaneous ($i=j$) and cooperative ($i \neq j$) decay processes due to near-field vacuum fluctuations. The third and fourth terms on the right-hand side of Eq.~(\ref{mainlindbladT}) correspond to thermally stimulated activated emission and absorption induced by near-field thermal fluctuations. Eq.~(\ref{mainlindbladT}) also shows that when cooperative decay rates $\Gamma_{ij} \, (i \neq j)$ are comparable to spontaneous decay rates $\Gamma_{ii}$, cooperative decay processes should not be neglected in EWJN effects. Including both spontaneous and cooperative decay rates in EWJN is one of the main contributions of this work. 
	
	\subsection{Spontaneous and cooperative decay rates: computational electromagnetics simulations}

    \begin{figure*}[t]
	    \centering
	    \includegraphics[width = 5 in]{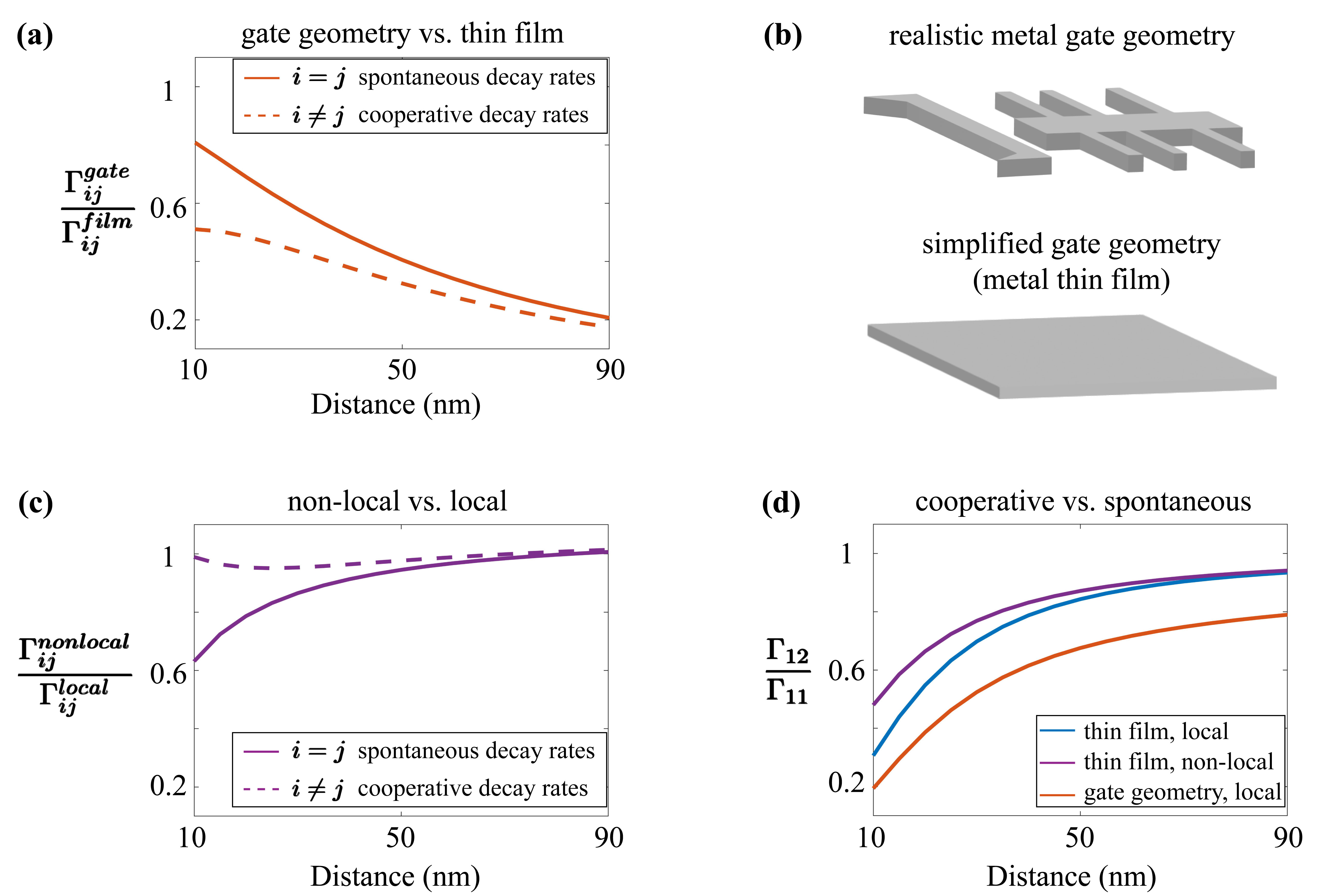}
	    \caption{Influence of metal gate geometries, non-locality, and cooperative decay on the decoherence of the two-spin-qubit system in the vicinity of metal gates. The two electron spin qubits with resonance frequencies $2\pi \times 39.33\,\mathrm{GHz}$ are separated by $50\, \mathrm{nm}$ and at the same distance (ranging from $10\,\mathrm{nm}$ to $90\,\mathrm{nm}$) from the metal gates. (a) Influence of metal gate geometries on the spontaneous and cooperative decay rates. Schematics of the thin film geometry and gate geometry are illustrated in (b). The metal gates are considered to be of aluminum with $100\,\mathrm{nm}$ thickness. Effects of finite gate geometry are prominent with increasing distance between qubits and metals. (b) Influence of the non-local dielectric response of metal gates on the spontaneous and cooperative decay rates. We employ the Lindhard formula to describe the non-local response. Non-local effects are essential for spontaneous decay rates when the qubits are very close to metal gates (distance $\lessapprox 10 \, \mathrm{nm}$) while having relatively less influence on cooperative decay rates. (c) Comparing the cooperative $\Gamma_{ij}(i\neq j)$ and spontaneous $\Gamma_{ij}(i = j)$ decay rates induced by near-field vacuum and thermal fluctuations in the vicinity of metal gates with the thin film geometry or realistic gate geometry and local or non-local dielectric responses. Compared with the spontaneous decay, cooperative decay becomes more prominent with increasing separation distance between qubits and metals.}
	    \label{fig:fig_DecayRates}
	\end{figure*}

    From macroscopic QED theory of EWJN (Appendix.~\ref{qedtheory}), the spontaneous ($i=j$) and cooperative ($i\neq j$) decay rates $\Gamma_{ij}$ are:
    
    \begin{align}
    \begin{aligned}\label{decayrates}
        \Gamma_{ij} &= \frac{2\mu_0}{\hbar} \mathbf{m}_{i} \cdot \big[ \nabla_{\mathbf{r}_i} \times \mathrm{Im} \overleftrightarrow{G}(\mathbf{r}_i,\mathbf{r}_j,\omega_+) \times \nabla_{\mathbf{r}_j} \big] \cdot \mathbf{m}_{j}^\dagger,\\
        &=\frac{2\mu_0}{\hbar} (\frac{\omega_{+}}{c})^2   \mathbf{m}_{i}  \cdot  \mathrm{Im} \, \overleftrightarrow{G}_m  (\mathbf{r}_i,\mathbf{r}_j,\omega_+)  \cdot \mathbf{m}_{j}^\dagger,
    \end{aligned}
    \end{align}
    where $\mu_0$ is the vacuum permeability, %$\omega_i$, $\mathbf{r}_{i}$, $\gamma_i$ and $\mathbf{m}_i=[\hbar \gamma_i / 2,-i \hbar \gamma_i / 2,0]$ are the resonance frequency, position, gyromagnetic ratio and spin magnetic moment of the $i \mathrm{th}$ qubit separately. 
    $\overleftrightarrow{G}(\mathbf{r}_i,\mathbf{r}_j,\omega_+)$ and $\overleftrightarrow{G}_m(\mathbf{r}_i,\mathbf{r}_j,\omega_+)$ represent the electric and magnetic dyadic Green's functions. 
	
    To solve the system dynamics governed by Eq.~(\ref{mainlindbladT}), we first need to calculate the EWJN-induced spontaneous and cooperative decay rates $\Gamma_{ij}$ (Eq.~(\ref{decayrates})). %, which is proportional to the imaginary part of magnetic dyadic Green's function $\overleftrightarrow{G_m}$. 
    Evanescent waves in conductive metallic materials greatly enhance fluctuations of EM fields, which are captured by $\mathrm{Im}\overleftrightarrow{G}_m$ \cite{FORD1984195}. Many factors, including positions of the spin qubits $\mathbf{r}_{i}$ and geometries and material properties of metallic contacts, determine $\mathrm{Im} \overleftrightarrow{G}_m$. Meanwhile, the host materials of solid-state spin qubits are generally non-magnetic, non-absorptive, or only weakly absorptive. In the microwave frequency range (relative to the resonance frequencies of most solid-state spin qubits), metals are much more dissipative than host materials ($\mathrm{Im} \, \varepsilon_{metal} \gg \mathrm{Im} \, \varepsilon_{host}$) and the total electromagnetic response is dominated by the imaginary part of metal response $\mathrm{Im} \, \varepsilon_{metal}$. Hence, contributions from host materials to the EM field fluctuations are negligible compared with contributions from highly dissipative metals~\cite{PhysRevLett.68.3698,kolkowitz2015johnsonnoise,Tenberg2019siliconrelaxation}. Although local-field corrections originating from local EM fields~\cite{kittel2018introduction, PhysRevApplied.18.044065} can be important for some fluctuation-induced interactions (e.g., van der Waals interactions and Casimir-Polder interactions~\cite{buhmann2013dispersion,Fiedler2017,PhysRevA.79.022903}), local-field effects on $\Gamma_{ij}$ of spin qubits embedded in nonmagnetic materials is expected to be small~\cite{PhysRevLett.74.880, PhysRevB.34.3700, PhysRevA.105.053704}. Considering the experimental results~\cite{PhysRevB.83.245123,kolkowitz2015johnsonnoise,Tenberg2019siliconrelaxation} show a good match with theory predictions of $\Gamma_{ij}$ of spin qubits embedded in dielectrics neglecting local-field effects, we will not consider local-field corrections in this paper. As a result, high-order corrections from host materials, including local field effects, can be neglected in $\mathrm{Im} \overleftrightarrow{G}_m$. In the following, we discuss $\mathrm{Im} \overleftrightarrow{G}_m$ calculations in two different scenarios: (a) simplified gate geometries in some proposed quantum processor architectures, where $\mathrm{Im} \overleftrightarrow{G}_m$ can be calculated analytically; (b) realistic metal gate geometries in experimentally relevant quantum computing devices, where one needs to employ advanced computational electromagnetics methods based on VIEs.
    
    In the first case, when the distance between spin qubits and metallic control systems is much smaller than the characteristic size of metallic control systems, as shown in some quantum network and processor architecture designs \cite{Li2018siliconnetwork,Veldhorst2017siliconcmosarchitecture,pezzagna2021quantum,Chen2020diamondprocessor}, one can simplify the actual metal geometry to a metal thin film. Here, at extremely low temperatures and when qubits are very close to metal gates, non-local effects of metals become important since electron-electron scatterings are dominant and amplify the effects of non-locality \cite{kolkowitz2015johnsonnoise}. Due to the translational symmetry of the metal thin film geometry, analytical expressions of $\mathrm{Im} \, \overleftrightarrow{G}_m$ exist, and we consider Lindhard theory of non-local dielectric function $\varepsilon(q,\omega)$ in the calculation (see Appendix. \ref{analyticalg} for equations and details) \cite{PhysRevB.89.115401,PhysRev.178.1201,FORD1984195}. 
    
    In the second case, we consider a realistic quantum computing device reported in a recent reference \cite{Huang2019fbenchmark}. Here, we model fluctuations of EM fields near the actual gate geometry in numerical simulations based on the volume integral equations (VIEs) method \cite{1984_SWG, jin2015theory}. Since the electromagnetic response is dominated by the imaginary part of metal permittivity~\cite{kolkowitz2015johnsonnoise}, we formulate a VIE in the metal gates and neglect the contribution from host materials to the EM field fluctuations. The VIE-based method offers great flexibility in modeling arbitrarily shaped metallic gates with tetrahedron-element-based discretization. It also captures conduction and displacement currents flowing along an arbitrary direction inside the metallic materials with the vector basis functions. In the simulations, we employ the SWG basis as the local vector basis function to expand the electromagnetic fields in the metal gates~\cite{1984_SWG}. Fast solvers \cite{OmarJiaoVIE:15, YifanWang_TAP2022} have also been developed to accelerate VIE computation. The relative error of our VIE simulations is controllable and estimated to be less than $5 \%$. (see Appendix. \ref{viemethod} for more details).

    Based on the simulations of $\Gamma_{ij}$, we discuss the effects of metal gate geometries and non-locality, and compare the cooperative and spontaneous decay rates induced by EWJN. Here, we consider two electron spin qubits separated by $50\,\mathrm{nm}$ at the same distance (ranging from 10 nm to 90 nm) away from the metal gates. The resonance frequencies of both qubits are $2\pi \times 39.33 \, \mathrm{GHz}$. In Fig.~\ref{fig:fig_DecayRates}(a), we demonstrate the influence of metal gate geometries on the decoherence of the two-spin-qubit system. We compare $\Gamma_{ij}$ of the two-spin-qubit system in the vicinity of metal gates with simplified thin film geometry and actual gate geometry (dimensions of the gate geometry in Appendix.~\ref{viemethod}). Schematics of the thin film geometry and gate geometry are illustrated in Fig.~\ref{fig:fig_DecayRates}(b). Conductivity $6.8\times 10^7 \, \mathrm{S/m}$ and thickness $100 \, \mathrm{nm}$ of the metal gates are the same in both calculations. It is observed that gate geometries can have a significant influence on EWJN when the distance between qubits and metals is comparable to the characteristic size of metallic control systems. Cooperative decay rates $\Gamma_{ij}\,(i \neq j)$ are more sensitive to gate geometries than spontaneous decay rates $\Gamma_{ij}\,(i = j)$. As a result, the thin film approximation for the gate geometry provides reasonable results when the metallic system has a characteristic size much larger than its distance from the spin qubits. 
    
    In Fig.~\ref{fig:fig_DecayRates}(c), we study the effects of the non-local dielectric response of the metal gates. We consider the thin film geometry with $100\, \mathrm{nm}$ thickness for this comparison. Non-locality is captured by the Lindhard non-local dielectric function with the plasma frequency $\omega_p=1.75 \times 10^{16}~\mathrm{Hz}$, Fermi velocity $v_f=2.02 \times 10^6~\mathrm{m/s}$, and electron collision frequency $\nu=3.65\times 10^{13}~\mathrm{Hz}$. Our results indicate that non-locality is essential for spontaneous decay rates when the qubits are very close to metal gates (distance $\lessapprox 10 \, \mathrm{nm}$) while having a relatively smaller contribution to cooperative decay rates. In Fig.~\ref{fig:fig_DecayRates}(d), we compare the cooperative and spontaneous decay rates induced by EWJN when we consider the metal gates to have the thin film or realistic gate geometry and local or non-local dielectric response. We find that the cooperative effects have significant contributions to the decoherence of the two-spin-qubit system when the distance between qubits and metals is comparable to the separation between the two qubits.
 
	\subsection{Gate infidelity $\Delta F$ induced by EWJN}

    \begin{figure*}[t]
	    \centering
	    \includegraphics[width = 6.8in]{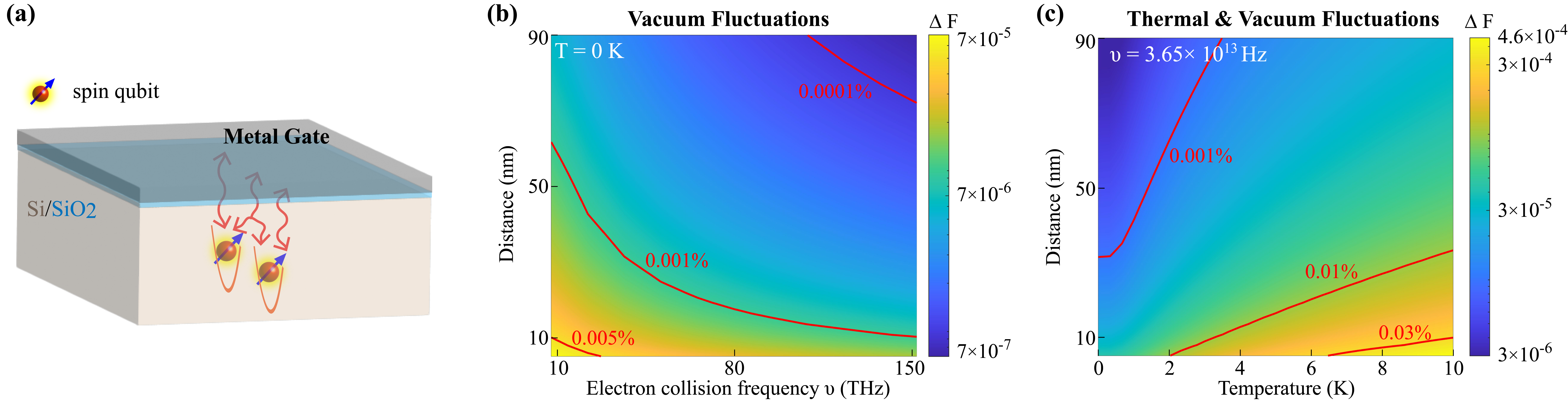}
	    \caption{Limits to CNOT gate fidelity from near-field vacuum and thermal fluctuations in silicon double quantum dot (DQD) system. (a) Schematic of the two-qubit system with simplified gate geometry in a quantum computing architecture based on silicon DQD. Red double-arrow lines represent all the spontaneous, stimulated, and cooperative decay processes of the two-qubit system induced by thermal and vacuum fluctuations. Top metal film (gray) represents the plane geometry of the metallic control system considered. %simplified aluminum lines and gates in the quantum processor.
	    %Hyperbolic potentials of the silicon DQD (orange) trap electron spin qubits.
	    Thin silica film is present between metal and silicon substrate. (b) CNOT gate infidelity $\Delta F $ due to vacuum fluctuations at $T=0~\mathrm{K}$ for different electron collision frequency $\nu$ and qubits’ distance $d$ from the metal film. (c) Dependence of $\Delta F $ on temperature $T$ and distance $d$ at fixed electron collision frequency $\nu$. Red lines in (b) and (c) represent constant contour lines of $\Delta F$.}
	    \label{fig:fig2}
	\end{figure*}
 
	With the calculated decay rates $\Gamma_{ij}$, quantum dynamics is simulated in the Liouville space (see Appendix.~\ref{dnumerical}). For a two-spin-qubit system with the initial density matrix $\rho_q(0)$, at time $t_f$ when the CNOT gate operation is finished, the actual final density matrix $\rho_q\, (\rho_q(0),t_f)$ can be calculated via Eq.~(\ref{mainlindbladT}). To this end, we define the CNOT gate fidelity as closeness between actual final density matrices $\rho_q\, (\rho_q(0),t_f)$ and theoretically ideal counterparts $\tilde{\rho}_q\,(\rho_q(0),t_f)$:
 
	\begin{equation}\label{equation5}
	    F=\sum_{\rho_q(0)} \mathrm{Tr} \, [\rho_q(\rho_q(0),t_f) \, \tilde{\rho}_q(\rho_q(0),t_f)]/4,
	\end{equation}
	where the initial density matrices $\rho_q(0)$ corresponding to the four computational basis states $|00\rangle,|01\rangle,|10\rangle,|11\rangle$ are considered in the average. Similar definitions have been employed in measuring and optimizing the performance of quantum gate operations and quantum algorithms~\cite{PhysRevApplied.17.024014,PhysRevLett.125.030501,PhysRevLett.124.220501,PhysRevLett.123.170503,Figgatt2017,PhysRevLett.114.200502,PhysRevApplied.19.024068}. Here, the influence of vacuum and thermal fluctuations on quantum gate operations is included in $\rho_q\, (\rho_q(0),t_f)$. Experimentally related parameters of MW control pulses are employed in our calculations of $\rho_q\, (\rho_q(0),t_f)$. Meanwhile, the theoretically ideal final density matrices $\tilde{\rho}_q\,(\rho_q(0),t_f)$ are obtained by applying perfect CNOT gate operations to $\rho_q(0)$. In this way, we can calculate the EWJN-affected CNOT gate fidelity $F$ with Eq.~(\ref{equation5}). By comparing the actual CNOT gate fidelity $F$ with the ideal counterpart $F_0$ where EWJN is ignored, we evaluate quantum gate infidelity $\Delta F= F_0-F$ induced by EWJN. An extended analysis of EWJN-induced average gate infidelity ($\Delta F$ averaged over all input states) is provided in Appendix~\ref{avgfidelity}.

	\subsection{Lindbladian engineering for pulse optimization}
	
	Finally, we present a Lindbladian engineering method that suppresses the influence of Markovian noise on two-spin-qubit quantum gate operations. In this part, we extend the noise source to include general Markovian noise induced dephasing where the Born-Markovian approximation is still justified. We also consider both the relaxation and dephasing processes of the spin qubit. We include a phenomenal Lindblad super-operator to describe the dephasing process of each single qubit $\hat{\hat{L}}_{\phi}$
	\begin{equation}\label{equation6}
	    \hat{\hat{L}}_{\phi} \rho_q(t)= \sum_i
		\frac{\Gamma^\phi_{i}}{2}[\hat{\sigma}_{z,i}\rho_q(t)\hat{\sigma}_{z,i}^{\dagger}-\rho_q(t)],
	\end{equation}
    where $\Gamma^\phi_{i}=\frac{1}{T^\phi_{i}}$ is the qubit pure dephasing rate at temperature $T$. $\hat{\sigma}_{z}$ is the Pauli matrix. System dynamics is then described by $d\rho_q(t) / dt= [\hat{H}(t),\rho_{q}(t)]/ i \hbar  + \hat{\hat{L}}_r \rho_{q}(t) + \hat{\hat{L}}_{\phi} \rho_{q}(t)$.
    
    We incorporate parameters of microwave pulses into a vector $\mathcal{X}$. $\mathcal{X}$ includes amplitudes, phases of microwave pulses, and lengths of pulses and delays. For a system with given $\{\Gamma_{ij},\Gamma^\phi_{i}\}$, through Eq.~(\ref{mainlindbladT})-(\ref{equation6}), we obtain a function $\mathcal{F}$ mapping pulse parameters $\mathcal{X}$ to quantum gate fidelity $F$:
    \begin{equation}\label{equation7}
        F=\mathcal{F}_{\{\Gamma_{ij},\Gamma^\phi_{i}\}}(\mathcal{X}).
    \end{equation}
	
	In Lindbladian engineering, for given $\{\Gamma_{ij},\Gamma^\phi_{i}\}$, we optimize $F$ by searching for optimal $\mathcal{X}$ in a constrained region. This is a nonlinear optimization problem with constraints. Here, this optimization is carried out using the interior-point method \cite{forsgren2002interior}. The interior-point method is widely used for inequality-constrained nonlinear optimization. It employs barrier functions to ensure that the final solutions are within the constraints~\cite{doi:10.1137/1.9781611970791}. The interior-point method has the advantage of handling inequality constraints and prevents the searching process from approaching the boundary of optimization too fast~\cite{GONDZIO2012587}. After the optimization, microwave control pulses with optimized parameters $\mathcal{X}$ realize quantum gate operations more robust against the near-field thermal and vacuum fluctuations. 
	
	\section{Influence of thermal and vacuum fluctuations in Two-Qubit Silicon Quantum Processor}\label{section3}
	
	In this section, we study CNOT gate infidelity $\Delta F$ induced by near-field thermal and vacuum fluctuations in the silicon double quantum dot (DQD) system. We employ experimental parameters of the two-spin-qubit systems \cite{Huang2019fbenchmark} in these simulations.
	
	The silicon DQD system consists of two electron spins trapped by hyperbolic potentials of quantum dots. The two spin qubits interact through exchange coupling, which dominates their unitary evolution. Both spontaneous and cooperative decay processes have important contributions to the non-unitary relaxation processes. The detailed representation of $\hat{H}_{ex}$, $\hat{H}_{mw}$, and $\hat{\hat{L}}_r$ in a rotating frame can be found in Appendix \ref{qedtheory}.
	
	\begin{figure*}
	    \centering
	    \includegraphics[width = 4.5in]{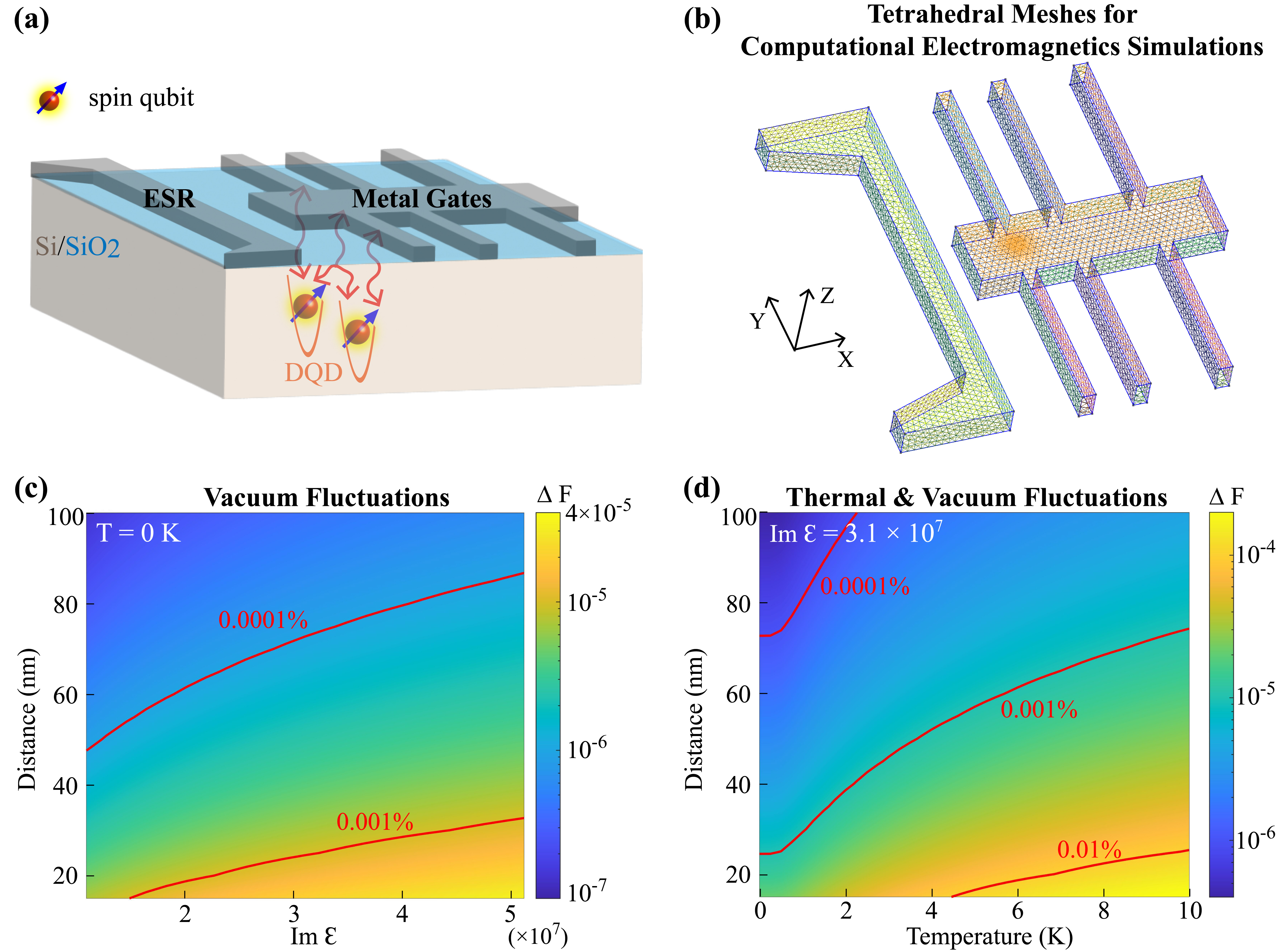}
	    \caption{Limits to CNOT gate fidelity from near-field vacuum and thermal fluctuations in a realistic quantum computing device based on silicon double quantum dot (DQD). (a) Schematic of a quantum computing device using silicon DQD electron spin qubits. Red double-arrow lines represent all the spontaneous, stimulated, and cooperative decay processes of the two-qubit system induced by thermal and vacuum fluctuations. Aluminum electron spin resonance (ESR) antennas and gates (gray) are on top of the silicon quantum dot. (b) Metal gate geometry with a tetrahedral mesh for computational electromagnetics simulations. Detailed dimensions of this gate geometry used in VIE simulations are depicted in Fig.~\ref{fig:apfig3} in Appendix \ref{viemethod}. (c) CNOT gate infidelity $\Delta F$ induced by vacuum fluctuations as a function of different metal dielectric properties $\mathrm{Im} \ \varepsilon$ and distance $d$ between metal gates and qubits. %$\Delta F$ is the infidelity caused by EWJN induced relaxation processes. 
	    (d) Dependence of $\Delta F $ on temperature $T$ and distance $d$ at fixed $\mathrm{Im} \, \varepsilon$. %Here, thermal and vacuum fluctuations contribute to EWJN effects. 
	    Red lines in (c) and (d) represent constant contour lines of $\Delta F$.}
	    \label{fig:fig3}
	\end{figure*}
	
	Figure~\ref{fig:fig2}(a) displays a schematic of the silicon DQD system. Here, considering the distance between spin qubits and metals is much smaller than the characteristic size of metals in some quantum processor architectures \cite{Li2018siliconnetwork,Veldhorst2017siliconcmosarchitecture}, we simplify the metal lines and gates to a metal film with $100~\mathrm{nm}$ thickness on top of the silicon substrate. CNOT gate operation is realized by the MW pulse generated from the metal electron spin resonance (ESR) stripline in the two-qubit processor. In the silicon DQD system, the operation temperature is around $30~\mathrm{mK}$ \cite{Huang2019fbenchmark}, and electron spins trapped in gate-defined quantum dots can be in close proximity to the metal gates. In this calculation, we assume that the two qubits are separated by $50\,\mathrm{nm}$ and at the same distance (between $5\,\mathrm{nm}$ and $90\,\mathrm{nm}$) from the simplified metal gates. As is shown in Fig.~\ref{fig:fig_DecayRates}(c, d), in this case, the non-local dielectric response of metals is significant at such low temperatures and small distances, and cooperative decay processes are crucial for the decoherence of the two-spin-qubit system. 
	Decay rates $\Gamma_{ij}$ are calculated based on the dyadic Green's function calculations presented in Appendix \ref{analyticalg}. 
    
	Aluminum is commonly used for building metal gates and antennas in quantum computing devices based on silicon DQD. Aluminum plasma frequency \mbox{$\omega_p=1.75 \times 10^{16}~\mathrm{Hz}$} and Fermi velocity \mbox{$v_f=2.02 \times 10^6~\mathrm{m/s}$} are constants in the range of temperature and material properties in our study \cite{smith1986intraband}. In the following, we first study the influence of EWJN induced by vacuum fluctuations on CNOT gate fidelity, then add thermal fluctuations and study the temperature dependence. We consider resonance frequencies \mbox{$\omega_i=\omega_j=2\pi \times 39.33~\mathrm{GHz}$} \cite{Huang2019fbenchmark} and spin magnetic dipole moments $\mathbf{m}_i=\mathbf{m}_j=[\frac{\hbar\gamma_e}{2},-i\frac{\hbar\gamma_e}{2},0]$ %, electron spin magenetic dipole moment electron gyromagnetic ratio \mbox{$\gamma_e=-1.76\times 10^{11} ~ \mathrm{rad / (s\cdot T)}$} 
    in our calculations.
    
    The electron collision frequency $\nu$ and spin qubits' distance $d$ from the metal film are crucial in determining vacuum fluctuations in the vicinity of aluminum. $\nu$ depends on material quality and fabrication process and is related to the non-local dielectric response of aluminum. In Fig.~\ref{fig:fig2}(b), we plot quantum gate infidelity $\Delta F$ induced by vacuum fluctuations for a range of $\{\nu, d\}$ values. Decreasing $d$ and $\nu$ will lead to an increasingly significant infidelity $\Delta F$ at $T=0 ~\mathrm{K}$. At small $d$ and low electron collision frequency $\nu$, maximum $\Delta F$ can exceed $0.005\%$. Considering the size of a logic qubit is sensitive to the fidelity of physical CNOT gates, this limits the minimum size and complexity of a practical silicon quantum processor with high-fidelity quantum logic operations. 
    
    In Fig.~\ref{fig:fig2}(c), we demonstrate the dependence of $\Delta F$ caused by EWJN on distance $d$ and environment temperature $T$. Both thermal and vacuum fluctuations contribute to EWJN in this case. Here, we keep electron collision frequency $\nu=3.65 \times 10^{13}~\mathrm{Hz}$ as a constant. $\Delta F$ increases with temperature because thermal fluctuations gradually dominate the decay processes of the two spin qubits. At $T=10~\mathrm{K}$, for quantum gate operations based on shallow quantum dots close to the aluminum film, gate infidelity $\Delta F$ induced by thermal and vacuum fluctuations can exceed $0.03\%$. This analysis shows that to build a processor based on silicon DQD with high-fidelity quantum logic operations and minimum size, it is important to mitigate the influence of EWJN to achieve higher physical gate fidelity. 
    
    \subsection*{Computational electromagnetics: realistic device geometries for silicon quantum dots}

    \begin{figure*}
	    \centering
	    \includegraphics[width = 6.8in]{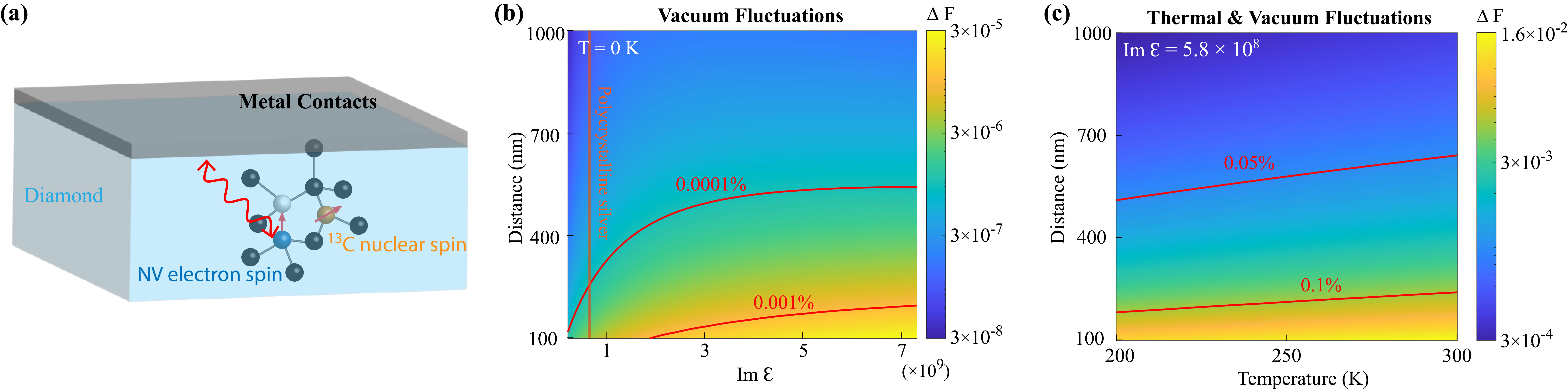}
	    \caption{Limits to CNOT gate fidelity from near-field vacuum and thermal fluctuations in the NV center system in diamond. (a) Schematic of an NV center system where electron spin and $^{13}\mathrm{C}$ nuclear spin are used as two qubits.  Top metal film (gray) represents the plane geometry of the metallic control system considered. (b) CNOT gate infidelity $\Delta F$ induced by vacuum fluctuations as a function of metal permittivity $\mathrm{Im} \ \varepsilon$ and distance $d$ between metal gates and qubits. (c) Dependence of $\Delta F$ on temperature of environment $T$ and distance $d$ at fixed $\mathrm{Im} \, \varepsilon$. Red lines in (b) and (c) represent constant contour lines of $\Delta F$.}
	    \label{fig:fig4}
	\end{figure*}
 
    In this subsection, we demonstrate the influence of EWJN on CNOT gate fidelity in a realistic silicon DQD device. 
    
    Figure~\ref{fig:fig3}(a) shows a schematic of the silicon DQD device considered in our model, which is obtained from the device reported in reference \cite{Huang2019fbenchmark}. From Fig.~\ref{fig:fig_DecayRates}(a), the thin film approximation is not valid for this gate geometry when the distance between qubits and metals is comparable to the characteristic size of metallic control systems. Here, we consider the thickness of the ESR antenna and metal gates to be $100 \, \mathrm{nm}$ and obtain accurate $\mathrm{Im} \, \overleftrightarrow{G}_m$ values using the VIE method (Appendix \ref{viemethod}). In Fig.~\ref{fig:fig3}(b), we illustrate the tetrahedron-element-based discretization used in the computational electromagnetics simulations. We assume that the two qubits are separated by 50 nm and at the same distance from the metal gates. The range of distance $d$ between electron spin qubits and metal gates in the simulations is from $15 \, \mathrm{nm}$ to $100 \, \mathrm{nm}$. As is shown in Fig.~\ref{fig:fig_DecayRates}(c), non-local effects only have a relatively small influence on the decoherence of the two-spin-qubit in this distance range. As a result, we neglect non-local effects in the VIE simulations. Detailed dimensions of this gate geometry and positions of the two qubits considered in VIE simulations are illustrated in Fig.~\ref{fig:apfig3} in Appendix \ref{viemethod}.
    
    We first study the influence of EWJN induced by vacuum fluctuations. In the local dielectric response regime, vacuum fluctuations of EM fields in the vicinity of aluminum gates are determined by distance $d$ and the imaginary part of aluminum permittivity $\mathrm{Im} \, \varepsilon$. $\mathrm{Im} \, \varepsilon$ is calculated from aluminum conductivity, which can be greatly influenced by the fabrication processes \cite{Tenberg2019siliconrelaxation}. We consider the range of aluminum conductivity to be from $2.4 \times 10^7 \, \mathrm{S/m}$ to $11.2 \times 10^7 \, \mathrm{S/m}$. We present the dependence of EWJN-induced CNOT gate infidelity $\Delta F$ on $\mathrm{Im} \, \varepsilon$ and $d$ in Fig.~\ref{fig:fig3}(c). It is clearly shown that $\Delta F$ increases with decreasing $d$ and increasing dissipation in metal gates. In Fig.~\ref{fig:fig3}(d), we consider the influence of both vacuum and thermal fluctuations. With increasing bath temperature, $\Delta F$ grows significantly and can exceed $0.01\%$ at $T>5 \, \mathrm{K}$.
 
	\section{Influence of thermal and vacuum fluctuations in two-qubit Diamond Quantum Processor}\label{section4}

    In this section, we evaluate the influence of EWJN on CNOT gate fidelity in the diamond NV center system. We follow the indirect control method and obtain parameters of the two-spin-qubit system from experiments~\cite{PhysRevLett.124.220501}.
	
    Figure~\ref{fig:fig4}(a) presents a schematic of the diamond NV center system. Electron spin (spin-1, control qubit) and one nearby $^{13}\mathrm{C}$ nuclear spin (spin-1/2, target qubit) form the two spin qubits for CNOT gate operations. Microwave control pulses generated by metallic antennas control the electron spin evolution. The $^{13}\mathrm{C}$ nuclear spin evolves under the hyperfine coupling between the two spin qubits. The microwave pulse sequence containing three pulses between four delays realizes CNOT gate operations. In the proposed scaling protocol \cite{Chen2020diamondprocessor, pezzagna2021quantum}, the characteristic size of metallic antennas and electrodes necessary for NV charge state and electron spin state control is much larger than their distance from the qubit system. As a result, to study the effects of thermal and vacuum fluctuations in a two-qubit diamond quantum processor, we can simplify these metal contacts to a metal film.  

    \begin{figure*}
	    \centering
	    \includegraphics[width = 6.5in]{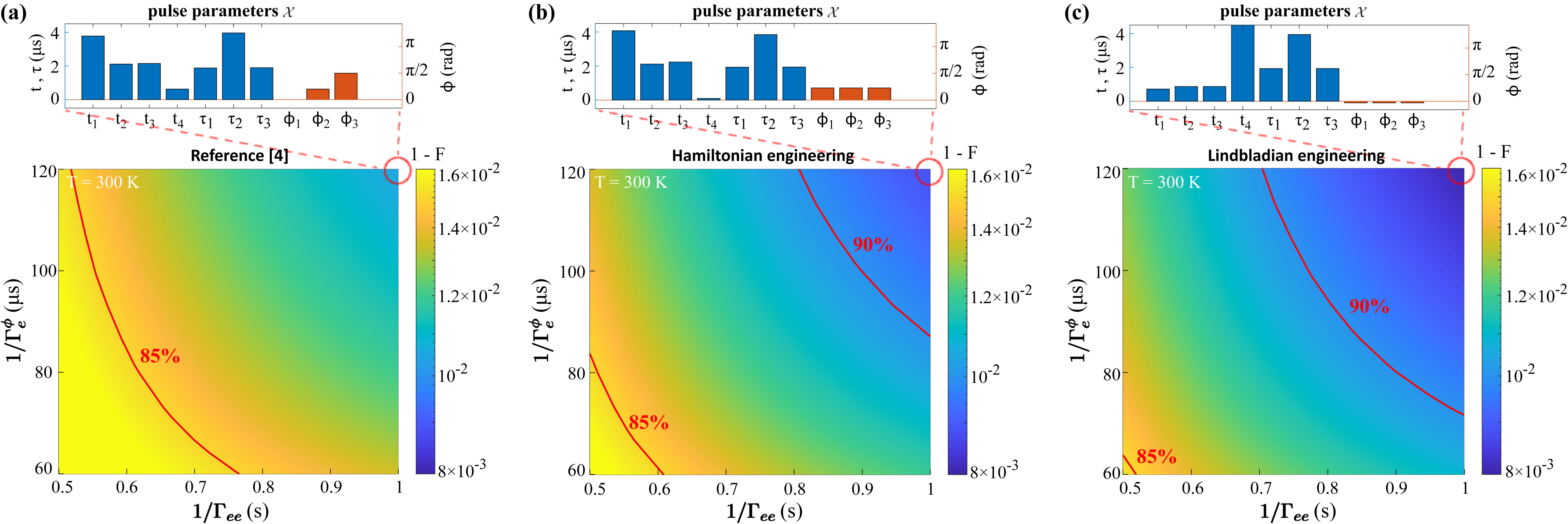}
	    \caption{Comparison of different MW pulse sequences and corresponding CNOT gate infidelity $1- F$ %caused by Markovian noise 
        in the diamond NV center system. Top bar plots represent the optimized pulse parameters, including lengths of four delays $t_1$, $t_2$, $t_3$, $t_4$ and phases and lengths of three pulses $\phi_1$, $\phi_2$, $\phi_3$, $\tau_1$, $\tau_2$, $\tau_3$. $\Gamma_{ee}$ and $\Gamma^\phi_e$ are the relaxation and dephasing rates of the electron spin qubit. (a) MW pulse sequence presented in reference~\cite{PhysRevLett.124.220501}. (b) MW pulse sequence optimized via Hamiltonian engineering. (c) MW pulse sequence optimized via Lindbladian engineering. Contour lines corresponding to gate fidelity $F=85\%$ and $F=90\%$ (out of the scope in (a)) are marked in red. MW pulse sequence optimized through Lindbladian engineering results in the highest achievable CNOT gate fidelity with every $\{\Gamma_{ee},\Gamma^\phi_{e}\}$.}
	    \label{fig:fig5}
	\end{figure*}
	
	Closed quantum system Hamiltonian $\hat{H}$ consists of two spin qubits' Hamiltonian $\hat{H}_{dd}$ and microwave Hamiltonian $\hat{H}_{mw}$. Here $\hat{H}_{dd}$ is dominated by dipole-dipole interaction. Since resonance frequency $\omega_e$ and gyromagnetic ratio $\gamma_e$ of the electron spin qubit are much larger than those of the $^{13}\mathrm{C}$ nuclear spin qubit \cite{yang2014electron,pezzagna2021quantum}, the spontaneous decay rate $\Gamma_{ee}$ of the electron spin qubit is much larger than other $\Gamma_{ij}$. As a result, the spontaneous and stimulated decay of the electron spin qubit will be dominant in the relaxation processes. We thus only consider terms related to the spontaneous and stimulated decay of the electron qubit in the Lindblad super-operator $\hat{\hat{L}}_r$. Detailed representations of $\hat{H}$ and $\hat{\hat{L}}_r$ capturing the dynamics in the truncated Hilbert space spanned by NV electron spin states $|0\rangle, |-1\rangle$ in the rotating frame are derived in Appendix \ref{qedtheory}.
	
	For the electron spin qubit with a distance $d$ greater than $100~\mathrm{nm}$ away from top metallic contacts, and gate operations around room temperature, we can ignore the non-local dielectric response of metallic contacts. We consider a silver film with a thickness of $300\,\mathrm{nm}$ as simplified metal gates. In the following, we first present CNOT gate infidelity $\Delta F$ due to EWJN induced by vacuum fluctuations. Next, we examine $\Delta F$ when thermal fluctuations are the dominant sources of EWJN. We consider the electron spin qubit resonance frequency $\omega_e=2\pi \times 2.458~\mathrm{GHz}$ and spin magnetic dipole moment $\mathbf{m}_e=[\frac{\sqrt{2}}{2}\hbar\gamma_e,i\frac{\sqrt{2}}{2}\hbar\gamma_e,0]$ %, $\gamma_e=-1.76\times 10^{11} ~ \mathrm{rad / (s\cdot T)}$ 
    in our calculations. Related pulse and system parameters are provided in Appendix \ref{qedtheory}.

	EWJN in the vicinity of silver is associated with the imaginary part of silver permittivity $\mathrm{Im} \, \varepsilon$. In Fig.~\ref{fig:fig4}(b), we present the dependence of CNOT gate infidelity $\Delta F$ on distance $d$ and silver permittivity $\varepsilon$ when vacuum fluctuations are the only sources of EWJN. The permittivity of poly-crystalline silver \cite{kolkowitz2015johnsonnoise} is marked by an orange vertical line in this plot. CNOT gate operations suffer from high $\Delta F$ when the electron spin is close to metallic contacts and metals have large $\mathrm{Im} \, \varepsilon$.
	
	In Fig.~\ref{fig:fig4}(c), we show the spatial and temperature dependence of $\Delta F$. Silver permittivity $\mathrm{Im} \, \varepsilon = 5.8\times 10^8$ is taken to be a constant in this calculation. The range of environment temperature considered is from $200~\mathrm{K}$ to $300~\mathrm{K}$, where thermal fluctuations are the major sources of EWJN. In this case, the maximum $\Delta F$ from EWJN exceeds $0.1 \% $ even when the electron spin qubit is over a distance of $100~\mathrm{nm}$ away from metallic contacts. This poses a limit to the minimum size and complexity of a practical diamond quantum processor with high-fidelity quantum logic operations.
 
	\section{Lindbladian Engineering to minimize influence of near-field thermal and vacuum fluctuations}\label{section5}
 
	In this section, we perform Lindbladian engineering in the two experimentally relevant systems to maximize quantum gate fidelity. We compare CNOT gate infidelity realized by MW pulses obtained through Lindbladian engineering and Hamiltonian engineering. Furthermore, we demonstrate that Lindbladian engineering can provide qubit driving protocols more robust against Markovian noise, including near-field thermal and vacuum fluctuations. 

	\subsection{Lindbladian engineering in two-qubit diamond quantum processor}\label{subsection1}
 
    \begin{figure*}
	    \centering
	    \includegraphics[width = 4.5in]{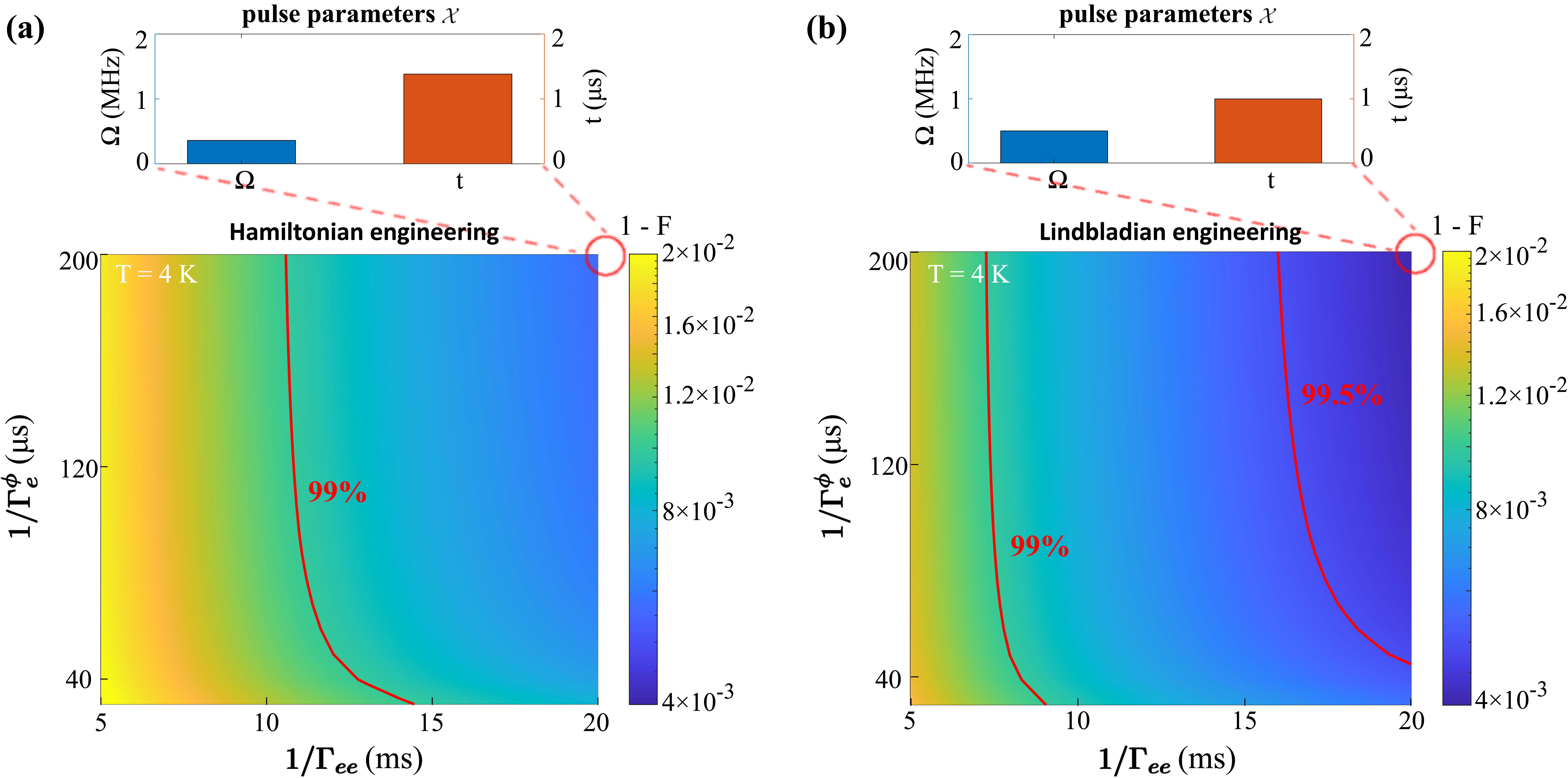}
	    \caption{Comparison of different MW pulse sequences and corresponding CNOT gate infidelity $1 - F$ in the silicon quantum dot system. (a) MW pulse optimized via Hamiltonian engineering. (b) MW pulse optimized via Lindbladian engineering. Top bar plots show different pulse parameters $\mathcal{X}$ in different cases. Contour lines corresponding to gate fidelity $F=99\%$ and $F=99.5\%$ (out of the scope in (a)) are marked in red. The high fidelity region of CNOT gate operations expands in the colormap corresponding to Lindbladian engineering.}
	    \label{fig:fig6}
	\end{figure*}
 
	Here, we consider Markovian relaxation and dephasing processes of the electron spin qubit in Lindbladian engineering. We examine the same qubit system shown in Sec.~\ref{section4} and consider an MW control pulse sequence consisting of three pulses within four delays. Pulse parameters include lengths of four delays $t_1,t_2,t_3,t_4$ and phases and lengths of three pulses $\phi_1,\phi_2,\phi_3$, $\tau_1,\tau_2,\tau_3$. Pulse parameter vector $\mathcal{X}$ is $\mathcal{X}=\{t_1,t_2,t_3,t_4,\tau_1,\tau_2,\tau_3,\phi_1,\phi_2,\phi_3 \}$.
	
	In the following, the relaxation and dephasing rates of the electron spin qubit are denoted as $\{\Gamma_{ee},\Gamma^\phi_e\}$, and we assume Markovian approximation is appropriate for describing spin decay and dephasing processes. CNOT gate is realized at room temperature $T=300~\mathrm{K}$. We investigate three different pulse sequences and their corresponding CNOT gate infidelity $1 - F$ %due to Markovian noise induced electron spin decay and dephasing
    , as is shown in Fig.~\ref{fig:fig5}. The three different pulse sequences are: original pulse sequence from \cite{PhysRevLett.124.220501} in Fig.~\ref{fig:fig5}(a), pulse sequence optimized via Hamiltonian engineering in Fig.~\ref{fig:fig5}(b), and pulse sequence optimized via Lindbladian engineering in Fig.~\ref{fig:fig5}(c). For $\{1/\Gamma_{ee},1/\Gamma^\phi_{e}\}=\{1~\mathrm{s}, 120~\mathrm{\mu s}\}$, ten parameters of the optimized control pulse sequence in the three different cases are shown in the bar plots of Fig.~\ref{fig:fig5}. In the three colormaps of CNOT gate infidelity, we find that the high fidelity region is the largest in Fig.~\ref{fig:fig5}(c) corresponding to Lindbladian engineering. This shows that it is possible to realize high-fidelity quantum gate operations in the range of large $\{\Gamma_{ee},\Gamma^\phi_{e}\}$. This is because control pulses obtained via Lindbladian engineering suppress the influence of Markovian noise on quantum gate fidelity. These results demonstrate that Lindbladian engineering can provide an optimal control protocol for this quantum computing system affected by near-field vacuum and thermal fluctuations.
	
    \subsection{Lindbladian Engineering in two-qubit silicon quantum processor}\label{subsection2}
 
	Here, we consider the Markovian relaxation and dephasing processes of the two-spin-qubit system in Lindbladian engineering. We examine the system presented in Sec.~\ref{section3} with a strong exchange interaction. Since only one microwave pulse is implemented, pulse parameters $\mathcal{X}$ consist of pulse length $t$ and $\Omega$ which is related to pulse strength.
	
	We use the interior-point method \cite{forsgren2002interior} to find optimal $\mathcal{X}$ through Hamiltonian engineering and Lindbladian engineering for given $\{\Gamma_{ee},\Gamma^\phi_{e}\}$, which represent the spontaneous decay and dephasing rates of both qubits. The cooperative decay rates are considered to be $\Gamma_{ee}/2$ in this optimization. Figure~\ref{fig:fig6} shows the optimized $\mathcal{X}$ and CNOT gate infidelity $1 - F$ induced by Markovian noise for the two cases. We observe the expansion of the high-fidelity region in the colormap associated with Lindbladian engineering. Limited by the small optimization space in the control protocol (only one microwave pulse), Lindbladian engineering only indicates that reducing pulse length can decrease the CNOT gate infidelity. 
	
	\section{Conclusion}\label{section6}
	
	In conclusion, we have combined macroscopic quantum electrodynamics theory, computational electromagnetics, and fluctuational electrodynamics to study the effects of near-field thermal and vacuum fluctuations on a two-spin-qubit quantum computing system. We examine limits to quantum gate fidelity from thermal and vacuum fluctuations in two experimentally relevant systems: diamond NV center and silicon quantum dot systems. We provide detailed calculations of CNOT gate infidelity due to EWJN as a function of distance, temperature, and dielectric properties of metallic contacts necessary for quantum gate operations. Although in this article, the gate fidelity is defined by averaging over the four computational basis states, the methods developed in this article can also be used to evaluate average gate fidelity (Appendix~\ref{avgfidelity}) and compare with the randomized benchmarking to study the average quantum gate infidelity induced by near-field electromagnetic field fluctuations~\cite{PRXQuantum.3.020357,RevModPhys.94.015004,PhysRevA.71.062310,nielsen2002simple} (Appendix~\ref{avgfidelity}). Even with the rapid progress of current technology, this influence is relatively less explored and can limit the minimum size and complexity of the spin-qubit-based quantum processor with high-fidelity quantum logic operations. Further, we propose Lindbladian engineering to mitigate the influence of EWJN on quantum gate fidelity, which can also suppress other Markovian noise impacts. We compare Hamiltonian engineering and Lindbladian engineering and demonstrate that control pulses optimized by Lindbladian engineering can realize higher CNOT gate fidelity by overcoming the effects of near-field thermal and vacuum fluctuations. Our findings help to reach the limits of two-spin-qubit quantum gate fidelity and accelerate the practical application of quantum computing.
	
	\section{Acknowledgements}\label{section7}
	This work was supported by the Defense Advanced Research Projects Agency (DARPA) under Applications Resulting from Recent Insights in Vacuum Engineering (ARRIVE) program and the Army Research Office under W911NF-21-1-0287.

	\appendix
	
    \section{Macroscopic Quantum Electrodynamics Theory of EWJN}\label{qedtheory}
    
    In this appendix, we employ the macroscopic quantum electrodynamics (QED) method to study the dynamics of two spin qubits driven by microwave control pulses in the presence of EWJN. Following the quantization framework in macroscopic QED \cite{scheel2009macroqed,scheel2009arxivqed,Yang_2020,cris2022fbound}, the total Hamiltonian can be written as:

    \begin{equation}
        \hat{H}_{tot}=\hat{H}_q+\hat{H}_f+\hat{H}_{int}+\hat{H}_{ex},
    \end{equation}
    
    \begin{equation}
        \hat{H}_q=\sum_{i=1}^2 \hbar \omega_{i} \, \hat{\sigma}_i^+ \hat{\sigma}_i^-,
    \end{equation}
    
    \begin{equation}
        \hat{H}_f=\int d^3 \mathbf{r} \int_0^\infty \hbar \omega \ \hat{\mathbf{f}}^\dagger (\mathbf{r},\omega) \hat{\mathbf{f}}(\mathbf{r},\omega),
    \end{equation}
    
    \begin{equation}
        \hat{H}_{int}=- \sum_{i=1}^2 (\mathbf{m}_{i}\hat{\sigma}_i^+ + \mathbf{m}_{i}^\dagger \hat{\sigma}_i^-) \cdot \hat{\mathbf{B}}(\mathbf{r}_i).
    \end{equation}
	
    Here, $\hat{H}_q$ is the Hamiltonian of the two spin qubits, $\hat{H}_f$ represents the Hamiltonian of the electromagnetic bath, $\hat{H}_{int}$ describes the interaction Hamiltonian between spin qubits and the electromagnetic bath, and $\hat{H}_{ex}$ represents the exchange coupling Hamiltonian between the exchange-coupled spin qubits. $\omega_i, \mathbf{m}_i, \mathbf{r}_i$ stand for the resonance frequency, spin magnetic dipole moment, and position of the $i \mathrm{th}\ (i=1,2)$ spin qubit. $\hat{\sigma}^{+(-)}_i=|1\rangle_i \langle0|_i \, (|0\rangle_i \langle1|_i)$ is the raising (lowering) operator of the $i \mathrm{th}$ spin qubit. $\hat{\mathbf{f}}^\dagger$ and $\hat{\mathbf{f}}$ are photon/polariton creation and annihilation operators satisfying the following commutation relations:
    
    \begin{equation}
        [\hat{\mathrm{f}}_\alpha(\mathbf{r},\omega), \,\hat{\mathrm{f}}_\beta^\dagger(\mathbf{r}',\omega')]=\delta_{\alpha\beta}\delta(\mathbf{r}-\mathbf{r'})\delta(\omega-\omega'), 
    \end{equation}
    \begin{equation}
        [\hat{\mathrm{f}}_\alpha(\mathbf{r},\omega), \,\hat{\mathrm{f}}_\beta(\mathbf{r}',\omega')]=0, 
    \end{equation}
    where $\alpha,\beta = x,y,z$, $\hat{\mathrm{f}}_\alpha$ is the $\alpha$ component of the vector $\hat{\mathbf{f}}$. The magnetic field operator $\hat{\mathbf{B}}(\mathbf{r})$ can be expressed in terms of $\hat{\mathbf{f}}^\dagger$, $\hat{\mathbf{f}}$ and the electric dyadic Green's function $\overleftrightarrow{G}(\mathbf{r},\mathbf{r}',\omega)$:
    
    \begin{equation}
        \hat{\mathbf{B}}(\mathbf{r}) = \int_0^\infty d\omega [\hat{\mathbf{B}}(\mathbf{r},\omega)+\hat{\mathbf{B}}^\dagger(\mathbf{r},\omega)],
    \end{equation}
    
    \begin{equation}
        \hat{\mathbf{B}}(\mathbf{r},\omega) = (i \omega)^{-1} \int d^3 \mathbf{r}' \ \nabla_\mathbf{r} \times \overleftrightarrow{G}(\mathbf{r},\mathbf{r}',\omega) \cdot \hat{\mathbf{f}}(\mathbf{r}',\omega).
    \end{equation}
    
    Exchange interaction Hamiltonian $\hat{H}_{ex}$ can be aproximated as \cite{Huang2019fbenchmark,Meunier2011effcphase}:
    
    \begin{equation}
        \hat{H}_{ex}=-J \ (|10\rangle \langle10| + |01\rangle \langle01|-|10\rangle \langle01| - |01\rangle \langle10|),
    \end{equation}
    where $J$ describes the strength of exchange interaction. In silicon quantum dot system, two spin qubits are coupled dominantly through $\hat{H}_{ex}$. In the NV center system, spin qubits are not coupled through the exchange interaction and we take $J=0$.
    
    In the following, we use the superscript $\mathrm{I}$ to denote operators and density matrices in the interaction picture with respect to $\hat{H}_0=\hat{H}_q+\hat{H}_f$. In the interaction picture, the Liouville–von Neumann equation is: 
    
    \begin{equation}\label{totliouville}
        \frac{d \rho^{\mathrm{I}}_{tot}(t)}{d t}=\frac{1}{i \hbar} [\hat{H}_{int}^\mathrm{I}(t)+\hat{H}_{ex}^\mathrm{I}(t),\rho^\mathrm{I}_{tot}(t)],
    \end{equation}
    where $\rho^{\mathrm{I}}_{tot}(t) = \rho^{\mathrm{I}}_{q}(t) \otimes \rho^{\mathrm{I}}_{f}(t)$ is the total density matrix, and $\rho^{\mathrm{I}}_{q}(t)$ and $\rho^{\mathrm{I}}_{f}(t)$ represent the density matrices of qubits and fields in the interaction picture separately. The integral form of Eq.~(\ref{totliouville}) is:
    \begin{equation}\label{integralform}
        \rho^{\mathrm{I}}_{tot}(t)=\rho^{\mathrm{I}}_{tot}(0)+\frac{1}{i \hbar} \int_0^t \, d\tau [\hat{H}_{int}^\mathrm{I}(\tau)+\hat{H}_{ex}^\mathrm{I}(\tau),\rho^\mathrm{I}_{tot}(\tau)].
    \end{equation}
    
    Substitute Eq.~(\ref{integralform}) back into Eq.~(\ref{totliouville}), we can obtain:
    
    \begin{multline}\label{A11}
        \frac{d \rho^{\mathrm{I}}_{tot}(t)}{d t} = \frac{1}{i \hbar} [\hat{H}_{int}^\mathrm{I}(t)+\hat{H}_{ex}^\mathrm{I}(t),\rho^\mathrm{I}_{tot}(0)]\\
        -\frac{1}{\hbar^2}\int_0^{t} d\tau \ [\hat{H}_{int}^\mathrm{I}(t)+\hat{H}_{ex}^\mathrm{I}(t),[\hat{H}_{int}^\mathrm{I}(\tau)+\hat{H}_{ex}^\mathrm{I}(\tau),\rho^\mathrm{I}_{tot}(\tau)]].
    \end{multline}
    
    The two-spin-qubit density matrix $\rho^{\mathrm{I}}_{q}(t)$ can be obtained by tracing out the field part: 
    
    \begin{widetext}
    \begin{align}
    \begin{aligned}\label{liouvilleeqaftertrace}
        \frac{d \rho^{\mathrm{I}}_{q}(t)}{d t} &= \frac{1}{i \hbar} [\hat{H}_{ex}^\mathrm{I}(t),\rho^\mathrm{I}_{q}(0)] -\frac{1}{\hbar^2}\int_0^{t} d\tau \ \mathrm{Tr}_f [\hat{H}_{int}^\mathrm{I}(t),[\hat{H}_{int}^\mathrm{I}(\tau),\rho^\mathrm{I}_{tot}(\tau)]]
        -\frac{1}{\hbar^2}\int_0^{t} d\tau \ [\hat{H}_{ex}^\mathrm{I}(t),[\hat{H}_{ex}^\mathrm{I}(\tau),\rho^\mathrm{I}_{q}(\tau)]]\\
        &=\frac{1}{i \hbar} [\hat{H}_{ex}^\mathrm{I}(t),\rho^\mathrm{I}_{q}(t)] -\frac{1}{\hbar^2}\int_0^{t} d\tau \ \mathrm{Tr}_f [\hat{H}_{int}^\mathrm{I}(t),[\hat{H}_{int}^\mathrm{I}(\tau),\rho^\mathrm{I}_{q}(\tau)]]\\
        &=\frac{1}{i \hbar} [\hat{H}_{ex}^\mathrm{I}(t),\rho^\mathrm{I}_{q}(t)] -\frac{1}{\hbar^2}\int_0^{t} d\tau \ \mathrm{Tr}_f [\hat{H}_{int}^\mathrm{I}(t) \hat{H}_{int}^\mathrm{I}(\tau) \rho^\mathrm{I}_{q}(\tau)-\hat{H}_{int}^\mathrm{I}(t) \rho^\mathrm{I}_{q}(\tau) \hat{H}_{int}^\mathrm{I}(\tau) \\
        & \qquad \qquad \qquad \qquad \qquad \qquad \qquad \qquad \qquad \qquad \qquad \qquad - \hat{H}_{int}^\mathrm{I}(\tau) \rho^\mathrm{I}_{q}(\tau) \hat{H}_{int}^\mathrm{I}(t) +  \rho^\mathrm{I}_{q}(\tau) \hat{H}_{int}^\mathrm{I}(\tau) \hat{H}_{int}^\mathrm{I}(t)].
    \end{aligned}
    \end{align}
    \end{widetext}
    
    It is clear that the first term in the last line of Eq.~(\ref{liouvilleeqaftertrace}) only contributes to the unitary evolution of $\rho^\mathrm{I}_{q}(t)$. The influence of fluctuating electromagnetic fields on $\rho^\mathrm{I}_{q}(t)$ is captured by the second term, where $\hat{H}_{int}^\mathrm{I}(t)$ is:

    \begin{widetext}
    \begin{multline}\label{interationHamiltonian}
        \hat{H}_{int}^\mathrm{I}(t) = -\sum_{j=1}^2 \bigl[ \int_0^\infty d\omega \ (i\omega)^{-1} 
        \int d^3\mathbf{r'} (\mathbf{m}_{i}\hat{\sigma}_i^+ e^{i\omega_i t} + \mathbf{m}_{i}^\dagger \hat{\sigma}_i^- e^{-i\omega_i t}) \cdot [\nabla_{\mathbf{r_i}} \times \overleftrightarrow{G}(\mathbf{r_i},\mathbf{r}',\omega)] \cdot 
        \hat{\mathbf{f}}(\mathbf{r}',\omega) e^{- i\omega t} + h.c. \bigr].
    \end{multline}
    \end{widetext}
    
    In the following, we employ the Born-Markovian approximation to simplify Eq.~(\ref{liouvilleeqaftertrace}). We assume that the influence of the two-spin-qubit system on the electromagnetic bath is small (Born approximation) \cite{breuer2002theory}. As a result, we have $\rho^{\mathrm{I}}_{tot}(t) = \rho^{\mathrm{I}}_{q}(t) \otimes \rho^{\mathrm{I}}_{f}(0)$ because $\rho^{\mathrm{I}}_{f}(t)$ is only negligibly affected by the two-spin-qubit system. We also assume that the bath correlation time $\tau_c$ is much smaller than the relaxation times of the system (Markovian approximation) \cite{breuer2002theory}. For two spin qubits with resonance frequencies $\omega_i, \, \omega_j$ satisfying  $|\omega_i-\omega_j| \ll \omega_i+\omega_j$, under the Markovian approximation, we have:
    
    \begin{widetext}
    \begin{subequations}\label{A15ab}
    \begin{align}
        \int_0^t ds \ e^{-i\omega (t-s)} e^{\pm i \omega_i t} e^{\mp i \omega_j s} &= e^{\pm 2 i \omega_- t}\int_0^\infty d\tau \ e^{- i (\omega \mp \omega_+) \tau} = e^{\pm 2 i \omega_- t} \ [\pi \delta(\omega \mp \omega_+) - i \mathcal{P} (\frac{1}{\omega \mp \omega_+})],\label{A14}\\
        \int_0^t ds \ e^{i\omega (t-s)} e^{\pm i \omega_i t} e^{\mp i \omega_j s} &= e^{\pm 2i\omega_-t}\int_0^\infty d\tau \ e^{i(\omega \pm \omega_+)\tau} =e^{\pm 2 i \omega_- t} \ [\pi \delta(\omega \pm \omega_+) + i \mathcal{P} (\frac{1}{\omega \pm \omega_+})]\label{A15},
    \end{align}
    \end{subequations}
    \end{widetext}
    where $\tau=t-s, \, \omega_-=(\omega_i-\omega_j)/2, \, \omega_+=(\omega_i+\omega_j)/2$.
    
    Substituting Eqs.~(\ref{A15ab}) into Eqs.~(\ref{liouvilleeqaftertrace}) and (\ref{interationHamiltonian}), we can obtain the open quantum system dynamics of the two-spin-qubit system. The relaxation processes within the computational subspace induced by near-field vacuum and thermal fluctuations of electromagnetic fields are captured by the trace-preserving Lindblad super-operator $\hat{\hat{L}}_{r}$. When the fluctuating vacuum fields ($\rho_{f}(t)=|0\rangle \langle0|$) interact with the two spin qubits, $\hat{\hat{L}}^\mathrm{I}_{r}$ describing the effects of near-field vacuum fluctuations on the two-spin-qubit system is:
    
    \begin{multline}
        \hat{\hat{L}}^\mathrm{I}_r \rho^\mathrm{I}_{q}(t) = \sum_{i,j} \Gamma_{ij}  [\hat{\sigma}_i^- \rho_q^{\mathrm{I}}(t) \hat{\sigma}_j^+ e^{-i(\omega_i-\omega_j)t}\\
        - \frac{1}{2} \rho_q^{\mathrm{I}}(t) \hat{\sigma}_i^+ \hat{\sigma}_j^- e^{i(\omega_i-\omega_j)t} - \frac{1}{2} \hat{\sigma}_i^+ \hat{\sigma}_j^- \rho_q^{\mathrm{I}}(t) e^{i(\omega_i-\omega_j)t}],
    \end{multline}
    where the spontaneous and cooperative decay rates $\Gamma_{ij}$ are:
    
    \begin{equation}
        \Gamma_{ij} = \frac{2 \mu_0}{\hbar} \mathbf{m}_{i} \cdot \big[ \nabla_{\mathbf{r}_i} \times \mathrm{Im} \overleftrightarrow{G}(\mathbf{r}_i,\mathbf{r}_j,\omega_+) \times \nabla_{\mathbf{r}_j} \big] \cdot \mathbf{m}_{j}^\dagger.
    \end{equation}
    
    Transforming back to the Schrodinger picture, we obtain the following Lindblad master equation:
    \begin{multline}\label{lindblad}
        \frac{d\rho_q(t)}{dt}=\frac{1}{i \hbar} [\hat{H}(t),\rho_{q}(t)] + \hat{\hat{L}}_r\rho_{q}(t) = \frac{1}{i \hbar} [\hat{H}(t),\rho_{q}(t)] \\ + \sum_{i,j} \Gamma_{ij}  [\hat{\sigma}_i^- \rho_q(t) \hat{\sigma}_j^+ 
        - \frac{1}{2} \rho_q(t) \hat{\sigma}_i^+ \hat{\sigma}_j^- - \frac{1}{2} \hat{\sigma}_i^+ \hat{\sigma}_j^- \rho_q(t)].
    \end{multline}
    
    The first term on the RHS of Eq.~(\ref{lindblad}) describes the unitary evolution of the two-spin-qubit system. Here, $\hat{H}$ consists of control Hamiltonians $\hat{H}_{mw}$ corresponding to microwave pulses and the coupling Hamiltonian between spin qubits. In the NV center system, the coupling Hamiltonian is dominated by $\hat{H}_{dd}$ which represents the dipole-dipole coupling between the NV electron spin and $^{13}\mathrm{C}$ nuclear spin. In the silicon dot system, the coupling Hamiltonian is dominated by $\hat{H}_{ex}$ which represents the exchange coupling between the two electron spins in silicon DQD. The second term on the RHS of Eq.~(\ref{lindblad}) describes the system relaxation process due to the coupling between the two-spin-qubit system and the electromagnetic bath. At finite temperature $T$, with the influence of thermal fluctuations included, we have:
    
    \begin{multline}\label{lindbladT}
        \frac{d\rho_q(t)}{dt}=\frac{1}{i \hbar} [\hat{H}(t),\rho_{q}(t)] + \hat{\hat{L}}_r\rho_{q}(t) = \frac{1}{i \hbar} [\hat{H}(t),\rho_{q}(t)] \\ + (\mathcal{N}+1) \sum_{i,j}  \Gamma_{ij}  [\hat{\sigma}_i^- \rho_q(t) \hat{\sigma}_j^+ 
        - \frac{1}{2} \rho_q(t) \hat{\sigma}_i^+ \hat{\sigma}_j^- - \frac{1}{2} \hat{\sigma}_i^+ \hat{\sigma}_j^- \rho_q(t)]\\ + \mathcal{N} \sum_{i,j} \Gamma_{ij}  [\hat{\sigma}_i^+ \rho_q(t) \hat{\sigma}_j^- 
        - \frac{1}{2} \rho_q(t) \hat{\sigma}_i^- \hat{\sigma}_j^+ - \frac{1}{2} \hat{\sigma}_i^- \hat{\sigma}_j^+ \rho_q(t)],
    \end{multline}
    where $\mathcal{N}$ is the mean photon number given by Eq.~(\ref{meanphotonnumber}).

    To this end, we simulate the dynamics of the two-spin-qubit system governed by Eq.~(\ref{lindbladT}) in the rotating frames. we use the superscript $\mathrm{R}$ to denote operators and density matrices in the rotating frame. Eq.~(\ref{lindbladT}) can be transformed into the rotating frame defined by the unitary operator $\hat{U}(t)$ through substituting $\hat{H}(t)$, $\sigma_i^\pm$, and  $\rho(t)$ with:
    \begin{equation*}
        \hat{H}^\mathrm{R}(t)=i \hbar \frac{d\hat{U}(t)}{dt} \hat{U}^{\dagger}(t)+\hat{U}(t)\hat{H}(t)\hat{U}^{\dagger}(t),
    \end{equation*}
    \begin{equation*}
        \sigma_i^{\mathrm{R},\pm}(t)=\hat{U}(t)\sigma_i^\pm(t)\hat{U}^{\dagger}(t),
    \end{equation*}
    \begin{equation*}
        \rho^\mathrm{R}(t)=\hat{U}(t)\rho(t)\hat{U}^{\dagger}(t).
    \end{equation*}
    
    \subsection{Rotating Frame for the Silicon Quantum Dot System}
    
    For the silicon quantum dot system, we simulate the dynamics of the two-spin-qubit system in a rotating frame defined by the following two unitary transformations sequentially \cite{Huang2019fbenchmark}:
    
    \begin{align*}
        \hat{U}^{(1)}(t)&=\begin{bmatrix}
            1 & 0 & 0 & 0 \\
            0 & \cos(\frac{\theta}{2}) & \sin(\frac{\theta}{2}) & 0 \\
            0 & -\sin(\frac{\theta}{2}) & \cos(\frac{\theta}{2}) & 0 \\
            0 & 0 & 0 & 1 
        \end{bmatrix},\\
        \hat{U}^{(2)}(t)&=\begin{bmatrix}
            e^{i\bar{E}_z t/
            \hbar} & 0 & 0 & 0 \\
            0 & e^{i (\delta \tilde{E}_z - J) t/2\hbar} & 0& 0 \\
            0 & 0 & e^{-i (\delta \tilde{E}_z + J) t/2\hbar} & 0 \\
            0 & 0 & 0 & e^{-i \bar{E}_z t/\hbar} 
        \end{bmatrix},
    \end{align*}
    where we employ experimental parameters $\theta=0.097, J/h=1.59 \, \mathrm{MHz}, \bar{E}_z/h=39.33 \, \mathrm{GHz}, \delta \tilde{E}_z/h = 13.35 \, \mathrm{MHz}$ of the two-spin-qubit system \cite{Huang2019fbenchmark}. 
    
    In the rotating frame, the unitary evolution part is governed by $\hat{H}^\mathrm{R} (t)$ that includes both the control pulse and exchange coupling Hamiltonians \cite{Huang2019fbenchmark}:
    
    \begin{widetext}
    \begin{align}
    \begin{aligned}
        \hat{H}^\mathrm{R}(t)&=\frac{\hbar}{2}\begin{pmatrix}0 & 0 & 0 & 0\\0 & 0 & 0 & \Omega\\0 & 0 & 0 & 0\\0 & \Omega^* & 0 & 0\end{pmatrix} + \frac{\hbar}{2} \frac{1-\tan(\frac{\theta}{2})}{1+\tan(\frac{\theta}{2})} \begin{pmatrix}0 & 0 & \Omega e^{iJt} & 0\\0 & 0 & 0 & 0\\\Omega^* e^{-iJt} & 0 & 0 & 0\\0 & 0 & 0 & 0\end{pmatrix}.
    \end{aligned}
    \end{align}
    \end{widetext}
    For CNOT gate operations in section \ref{section3}, we consider a single microwave pulse with $\Omega=2\pi\times 0.41 \, \mathrm{MHz}$ and length $1.2195\,\mathrm{\mu s}$.
    In the rotating frame, the non-unitary evolution part is captured by $\hat{\hat{L}}^\mathrm{R}_r$, which can be obtained by substituting $\sigma_i^\pm$ in $\hat{\hat{L}}_r$ with $\sigma_i^{\mathrm{R},\pm}(t)$:
    
    \begin{widetext}
    \begin{subequations}
    \begin{align}
        \sigma_1^{\mathrm{R},-}(t) & = e^{-i \bar{E}_z t/\hbar}\begin{pmatrix}0 & 0 & 0 & 0 \\ e^{i (\delta \tilde{E}_z - J) t/2\hbar}\sin(\frac{\theta}{2}) & 0 & 0 & 0 \\ e^{-i (\delta \tilde{E}_z + J) t/2\hbar} \cos(\frac{\theta}{2}) & 0 & 0 & 0 \\ 0 & e^{-i (\delta \tilde{E}_z - J) t/2\hbar} \cos(\frac{\theta}{2}) & -e^{i (\delta \tilde{E}_z + J) t/2\hbar}\sin(\frac{\theta}{2}) & 0 \end{pmatrix}, \\
        \sigma_2^{\mathrm{R},-}(t) & = e^{-i \bar{E}_z t/\hbar} \begin{pmatrix}0 & 0 & 0 & 0 \\ e^{i (\delta \tilde{E}_z - J) t/2\hbar}\cos(\frac{\theta}{2}) & 0 & 0 & 0 \\ - e^{-i (\delta \tilde{E}_z + J) t/2\hbar} \sin(\frac{\theta}{2}) & 0 & 0 & 0 \\ 0 & e^{-i (\delta \tilde{E}_z - J) t/2\hbar} \sin(\frac{\theta}{2}) & e^{i (\delta \tilde{E}_z + J) t/2\hbar}\cos(\frac{\theta}{2}) & 0 \end{pmatrix}.
    \end{align}    
    \end{subequations}
    
    \end{widetext}

    \subsection{Rotating Frame for the NV Center System}
    For the NV center system, we consider a rotating frame defined by the following unitary operator:
    
    \begin{equation*}
        \hat{U}(t) = e^{i \omega t \hat{s}_{z,e}}.
    \end{equation*}
    
    Here, $\omega$ is the resonance frequency of the electron spin qubit, $\hat{s}_{z,e}$ is the $z$ component of Pauli matrix in the truncated Hilbert space corresponding to electron spin qubit.
    
    The unitary evolution part is governed by $\hat{H}_{dd}^\mathrm{R} (t)$ and $\hat{H}_{mw}^\mathrm{R} (t)$ \cite{PhysRevLett.124.220501}:
    
    \begin{multline}
        \hat{H}^R_{dd}(t) = \hbar |0\rangle \langle0| \otimes [-\nu_c I_z] \\
        + \hbar|-1\rangle \langle-1| \otimes [-(\nu_c+A_{zz}) I_z-A_{zx}I_x],
    \end{multline}
    \begin{equation}
        \hat{H}^R_{mw}(t)=\hbar \omega_1[cos(\phi_i)s_x+sin(\phi_i)s_y],
    \end{equation}
    where $\nu_c=2\pi\times 0.158\, \mathrm{MHz}, \, A_{zz}=-2\pi\times0.152\, \mathrm{MHz}, \, A_{zx}=2\pi\times0.110\, \mathrm{MHz}, \, \omega_1=2\pi\times0.5 \, \mathrm{MHz}$ \cite{PhysRevLett.124.220501}. $\phi_i$ represents the phase of $i\mathrm{th}$ pulse. The control pulse sequence for CNOT gate operations considered in section \ref{section4} is presented in the bar plot in Fig. \ref{fig:fig5}(a). 
    
    Here, since $\sigma_i^{R,\pm}(t)=\hat{U}(t)\sigma_i^{\pm}\hat{U}^{\dagger}(t)=e^{\pm i\omega t} \,\sigma_i^\pm$, Lindblad super-operator $\hat{\hat{L}}^\mathrm{R}_r$ in the rotating frame will have the same form as $\hat{\hat{L}}_r$ in the lab frame.
    
	\section{Magnetic Dyadic Green's Function}\label{analyticalg}
	
	As is shown in section~\ref{section2}, to study the effects of near-field vacuum and thermal fluctuations on quantum gate fidelity, it is important to evaluate the magnetic dyadic Green's function $\overleftrightarrow{G}_m(\mathbf{r}_i,\mathbf{r}_j,\omega)$ near metal gates. In this appendix, we present the analytical expressions of magnetic dyadic Green's function $\overleftrightarrow{G}_m(\mathbf{r}_i,\mathbf{r}_j,\omega)$ in the vicinity of metal gates with the thin film geometry. We use $\hat{\mathbf{x}}, \hat{\mathbf{y}}, \hat{\mathbf{z}}$ to denote the unit vectors in the $x,y,z$ direction. Without loss of generality, we assume the metal thin film is perpendicular to the $\hat{\mathbf{z}}$ direction.
	
	The general electric dyadic Green's function $\overleftrightarrow{G}(\mathbf{r}_i,\mathbf{r}_j,\omega)$ is defined by the following equation~\cite{novotny_hecht_2012}:
	
	\begin{equation}\label{dyadicdef}
	    \nabla \times\nabla \times \overleftrightarrow{G} (\mathbf{r_{i},r_{j}},\omega) - k_0 ^2 \overleftrightarrow{G} (\mathbf{r_{i},r_{j}},\omega) = \overleftrightarrow{I} \delta(\mathbf{r_{i}}-\mathbf{r_{j}}),
	\end{equation}
	where $k_0=\frac{\omega}{c}$ is the free-space wave vector and $\overleftrightarrow{I}$ is the $3\times3$ identity matrix.
	
	The solution to Eq.~(\ref{dyadicdef}) can be expressed in terms of incident and reflected fields as $\overleftrightarrow{G}=\overleftrightarrow{G}_0 + \overleftrightarrow{G}_{ref}$. $\overleftrightarrow{G}_0$ is the free-space electric dyadic Green's function. %Since evanescent waves in metals greatly enhance the has a negligible contribution to the $\Gamma_{ij}$ of spin qubits in the vicinity of metals gates. 
    The total electromagnetic response is dominated by the reflected part $\overleftrightarrow{G}_{ref}$. $\overleftrightarrow{G}_{ref}(\mathbf{r}_i,\mathbf{r}_i,\omega)$ in the vicinity of a metal film is \cite{novotny_hecht_2012}:
	\begin{multline}
	    \overleftrightarrow{G}_{ref} (\mathbf{r_{i},r_{i}},\omega)=\frac{i}{8\pi k_0^2} \int_0^\infty \frac{qdq}{k_z}e^{2 i k_z d}	 \\   [(k_0^2 r_s- k_z^2 r_p)(\hat{\mathbf{x}}\otimes\hat{\mathbf{x}}+\hat{\mathbf{y}}\otimes\hat{\mathbf{y}})+2 q^2 r_p \hat{\mathbf{z}}\otimes\hat{\mathbf{z}}],
	\end{multline}
	where $q=|\mathbf{q}|$, $\mathbf{q}$ is the component of the wavevector parallel to the metal film, 
	$d=\mathbf{r_{i}} \cdot \hat{\mathbf z}$, $k_z = \sqrt{k_0^2-q^2}$ is the z component of the wavevector perpendicular to the metal film, $r_s$ and $r_p$ are the reflection coefficients for the s- and p- polarized light.

	From Eq.~(\ref{decayrates}), the magnetic dyadic Green's functions $\overleftrightarrow{G}_m(\mathbf{r_{i},r_{j}},\omega)$ are defined as:
    \begin{multline}
         \overleftrightarrow{G}_m (\mathbf{r_{i},r_{j}},\omega)= \frac{1}{k_0 ^2} \nabla_i \times \overleftrightarrow{G}(\mathbf{r_{i},r_{j}},\omega) \times \nabla_j.
    \end{multline}

    Using the Levi-Civita symbols $\epsilon_{\alpha kl}$ and $\epsilon_{\beta nm}$, components $\left[\overleftrightarrow{G}_m\right]_{\alpha \beta}$ of the magnetic dyadic Green's functions can be expressed as:
    \begin{equation}
        \left[\overleftrightarrow{G}_m (\mathbf{r_{i},r_{j}},\omega)\right]_{\alpha \beta}=\epsilon_{\alpha kl}\epsilon_{\beta nm}\nabla_i^k\nabla_j^n \left[\overleftrightarrow{G}(\mathbf{r_{i},r_{j}},\omega)\right]^{lm}.
    \end{equation}

    Similar to the electric dyadic Green's functions, $\overleftrightarrow{G}_m$ can be decomposed into the free-space part $\overleftrightarrow{G}_{m,0}$ and reflected part $\overleftrightarrow{G}_{m,ref}$. Due to the existence of metal gates, the free-space part has a negligible contribution to $\mathrm{Im} \overleftrightarrow{G}_m$ in the near-field. Hence, $\mathrm{Im} \overleftrightarrow{G}_m \approx \mathrm{Im} \overleftrightarrow{G}_{m,ref}$. In the following, we neglect the contribution from $\overleftrightarrow{G}_{m,0}$ and do not distinguish the differences between $\overleftrightarrow{G}_m$ and $\overleftrightarrow{G}_{m,ref}$ since only the imaginary part of $\overleftrightarrow{G}_m$ is important for our calculations of $\Gamma_{ij}$.

    The reflected part of the magnetic dyadic Green's function $\overleftrightarrow{G}_{m}$ (we have dropped the subscript $ref$) in the vicinity of a metal film is: 
	\begin{widetext}
    \begin{subequations}\label{anaG}
    \begin{equation}\label{anaspd}
	        \overleftrightarrow{G}_m (\mathbf{r_{i},r_{i}},\omega)
	        =\frac{i}{8\pi k_0^2} \int_0^\infty \frac{qdq}{k_z}e^{2 i k_z d}
	        [(k_0^2 r_p- k_z^2 r_s)(\hat{\mathbf{x}}\otimes\hat{\mathbf{x}}+\hat{\mathbf{y}}\otimes\hat{\mathbf{y}})\\
	        +2 q^2 r_s \hat{\mathbf{z}}\otimes\hat{\mathbf{z}}],
	\end{equation}	
	\begin{equation}\label{anacpd}
        \overleftrightarrow{G}_m (\mathbf{r}_i,\mathbf{r}_j,\omega)=\frac{i}{8 \pi^2}\int \frac{d \mathbf{q}}{k_z} e^{i \mathbf{q}(\mathbf{r}_i-\mathbf{r}_j)} e^{i k_z(z_i+z_j)} \biggl( \frac{r_p}{q^2}\begin{bmatrix} q_y^2&-q_x q_y&0\\-q_x q_y&q_x^2&0\\0&0&0 \end{bmatrix}+ \frac{r_s}{k_0^2q^2}\begin{bmatrix} -q_x^2 k_z^2 & -q_x q_y k_z^2 & -q_x k_z q^2\\-q_x q_y k_z^2 & -q_y^2 k_z^2 & -q_y k_z q^2\\q^2 q_x k_z & q^2 q_y k_z & q^4 \end{bmatrix} \biggr),
    \end{equation}
    \end{subequations}
	\end{widetext}
	where $\mathbf{q}$ is the component of the wavevector parallel to the metal film, $q_x=\mathbf q \cdot \hat{\mathbf x}$, $q_y=\mathbf q \cdot \hat{\mathbf y}$. $z_i$ and $z_j$ are the z components of $\mathbf{r}_i$ and $\mathbf{r}_j$. Here, $\overleftrightarrow{G}_m (\mathbf{r_{i},r_{i}},\omega)$ is related to the spontaneous and stimulated decay rates of a spin qubit, while $\overleftrightarrow{G}_m (\mathbf{r_{i},r_{j}},\omega)$ is related to the cooperative decay rates of the two-spin-qubit system.
	
	When the non-local dielectric response is neglected, for a non-magnetic metal film with thickness of $t$, $r_s$ and $r_p$ in Eq.~(\ref{anaG}) are given by the Fresnel reflection coefficients:
	\begin{equation}
	    r_s(q)=\frac{k_z^2-k_{zm}^2}{k_z^2+k_{zm}^2 + 2 i k_z k_{zm}\cot{(k_{zm} t)}},
	\end{equation}
	
	\begin{equation}
	    r_p(q)=\frac{\varepsilon^2 k_z^2-k_{zm}^2}{\varepsilon^2 k_z^2+k_{zm}^2+2 i \varepsilon k_z k_{zm} \cot{(k_{zm} t)}},
	\end{equation}
	where $\varepsilon=\varepsilon(\omega)$ is the permittivity of metal contacts, $k_{zm}=\sqrt{\varepsilon k_0^2 - q^2}$.
	
	Non-local dielectric response of metals can be captured by the Lindhard model \cite{FORD1984195,10013390197}. Reflection coefficients $r_s$ and $r_p$ for a metal thin film with thickness of $t$ and Lindhard non-local permittivity $\varepsilon(\omega, k)$ are \cite{PhysRev.178.1201,PhysRevB.89.115401}:
	
	\begin{equation}
	    r_s(q)=\frac{1}{2}\sum_{i=e,o}\frac{\frac{2ik_z}{t k_0^2}\sum_{n=-\infty}^{\infty} \frac{1}{\varepsilon_t(\omega,k_{i,n})-\frac{k_{i,n}^2}{k_0^2}}-1}{\frac{2ik_z}{t k_0^2}\sum_{n=-\infty}^{\infty}\frac{1}{\varepsilon_t(\omega,k_{i,n})-\frac{k_{i,n}^2}{k_0^2}}+1},
	\end{equation}
	
	\begin{equation}
	    r_p(q)=\frac{1}{2}\sum_{i=e,o}\frac{1-\frac{2q}{t}\Sigma_{n=-\infty}^{\infty} \frac{1}{k_{i,n}^2 \varepsilon_l(\omega,k_{i,n})}}{1+\frac{2q}{t}\Sigma_{n=-\infty}^{\infty} \frac{1}{k_{i,n}^2 \varepsilon_l(\omega,k_{i,n})}},
	\end{equation}
	where $k_{i,n}^2=q^2+p_{i,n}^2, p_{e,n}=\frac{2n\pi}{t}, p_{o,n}=\frac{(2n+1)\pi}{t}$. 
	
	Longitudinal and transverse Lindhard non-local permittivity $\varepsilon(\omega, k)$ are \cite{10013390197,FORD1984195}:

	\begin{equation}
	    \varepsilon_l(\omega,k)= 1 + \frac{3\omega_p^2}{\omega + i \nu} \frac{u^2 F_l(u)}{\omega + i \nu F_l(u)},
	\end{equation}
	
	\begin{equation}
	    \varepsilon_t(\omega,k)= 1- \frac{\omega_p^2}{\omega(\omega + i \nu)} F_t(u),
	\end{equation}

	\begin{equation}
	    u=\frac{\omega+i \nu}{k v_F},
	\end{equation}
	
	\begin{equation}
	    F_l(u)=1-\frac{1}{2}u \ln \frac{u+1}{u-1},
	\end{equation}
	
	\begin{equation}
	    F_t(u)=\frac{3}{2} u^2 -\frac{3}{4} u (u^2-1) \ln \frac{u+1}{u-1},
	\end{equation}
	where $\omega_p$ is the plasma frequency, $\nu$ is the electron collision frequency, $v_F$ is the Fermi velocity.
	
	\section{Computational Electromagnetics Simulations of Magnetic Dyadic Green's Function}\label{viemethod}
	
	In this appendix, we present the computational electromagnetics simulations of magnetic dyadic Green's function $\overleftrightarrow{G}_m$ in the vicinity of metal gates with arbitrary geometry. Similar to Appendix~\ref{analyticalg}, in the following, we only consider the scattered part $\overleftrightarrow{G}_{m,ref}$ and drop the subscript $ref$. Due to the lack of translational symmetry of metal gates, $r_s$ and $r_p$ become ill-defined, and approaches to calculate $\overleftrightarrow{G}_m$ in Appendix \ref{analyticalg} are no longer applicable. Here, we discuss how to obtain $\overleftrightarrow{G}_m$ close to metal gates in a quantum computing device via the volume integral equations (VIEs) method. Perturbative methods, including the Born-series expansion, have also been proposed to solve the scattering Green's function~\cite{gbur_2011,PhysRevA.89.062512,PhysRevA.92.022503,PhysRevResearch.2.013308}. However, considering that the Born-series-based iteration fails to converge for high contrast ratio~\cite{Kleinman:90}, in this article, we employ the numerical method to directly solve the VIEs without using the approximate Born-series-based iteration.
	
	\subsection{Magnetic dyadic Green's function}
	
	The magnetic field at position $\mathbf{r}$ generated by current density $\mathbf{j}(\mathbf{r'},\omega)$ is:
	
    \begin{equation}\label{magneticf}
        \mathbf{H}(\mathbf{r},\omega)=\int_V[\nabla \times \overleftrightarrow{\mathbf{G}}(\mathbf{r},\mathbf{r'},\omega)] \, \mathbf{j}(\mathbf{r'},\omega)dV'.
    \end{equation}
    
    A point magnetic dipole at the position $\mathbf{r_0}$ with dipole moment $\mathbf{m}$ and oscillating frequency $\omega$ is equivalent to a closed electric current loop with current density:
    
    \begin{equation}\label{magneticcd}
        \mathbf{j}(\mathbf{r'},\omega)=-\mathbf{m}(\mathbf{r}_0,\omega) \times \nabla^{'} \delta (\mathbf{r'}-\mathbf{r}_0).
    \end{equation}
    
    Here, $\delta(\mathbf{r})$ represents the Dirac delta function. Substitute Eq.~(\ref{magneticcd}) into Eq.~(\ref{magneticf}):
    
    \begin{align}
    \begin{aligned}\label{magneticf2}
        &\mathbf{H}(\mathbf{r},\omega)\\
        &=\int_V[\nabla \times \overleftrightarrow{\mathbf{G}}(\mathbf{r},\mathbf{r'},\omega)]\cdot [-\mathbf{m} \times \nabla^{'} \delta (\mathbf{r'}-\mathbf{r_0})]dV'\\
        &=\int_V[\nabla \times \overleftrightarrow{\mathbf{G}}(\mathbf{r},\mathbf{r'},\omega) \times \nabla^{'}]  \cdot \mathbf{m} \, \delta (\mathbf{r'}-\mathbf{r_0})dV'\\
        &=k_0^2 \, \overleftrightarrow{G}_m (\mathbf{r,r_0},\omega) \cdot \mathbf{m}(\mathbf{r}_0,\omega),
    \end{aligned}
    \end{align}
    where we have used the following relation to simplify Eq.~(\ref{magneticf2}):
    
    \begin{equation}
        \int dx \, f(x) \delta'(x)= -\int dx\, \delta (x) f'(x),
    \end{equation}
    where $\delta'(x)$ and $f'(x)$ represents the first derivatives of $\delta(x)$ and $f(x)$ with respect to $x$. From Eq.~(\ref{magneticf2}), to obtain $\overleftrightarrow{G}_m (\mathbf{r,r_0},\omega)$, we can place a test point magnetic dipole at $\mathbf{r_0}$ near the metallic gate structure (the scatterer), and use volume integral equations (VIE) to solve the scattered magnetic field at $\mathbf{r}$. By repeating this procedure three times with magnetic dipoles oriented along X, Y, and Z axes separately, we can obtain all the nine components of the magnetic dyadic Green's function $\overleftrightarrow{G}_m (\mathbf{r,r_0},\omega)$.

    \subsection{VIE Formulation}
    Since the size of the scatterer (metal gates) is comparable to or smaller than the skin depth of the material and the test dipole is placed very close, we need to solve the field inside the metal gates in order to get the scattered field. Here we use the VIE method, which is robust against the low-frequency breakdown. 

    In the simulation, we treat the scatterer (metal gates) as a dissipative medium with permittivity $ \varepsilon(\mathbf{r}) = \varepsilon_0(\varepsilon_r  - \frac{\sigma}{\omega \epsilon_0}i) $, occupying the volume $ V $ in the space. Consider the electric field at $\mathbf{r} \in V$, $ \mathbf{E}(\mathbf{r}) =  \frac{\mathbf{D}(\mathbf{r})}{\varepsilon(\mathbf{r})} = \mathbf{E}^i + \mathbf{E}^{sc} $, where $ \mathbf{E}^i $ and $\mathbf{E}^{sc} $ denote the incident and scattered field respectively. Therefore we can formulate an integral equation with an unknown $\mathbf{D}$-field as \cite{jin2015theory}:
    \begin{multline}\label{vie}
        \mathbf{E}^i(\mathbf{r}) = \mathbf{E}(\mathbf{r}) - \mathbf{E}^{sc}(\mathbf{r}) = \frac{\mathbf{D}(\mathbf{r})}{\varepsilon(\mathbf{r})} - \int_V \Big \{ \mu_0 \omega^2 \kappa (\mathbf{r'}) \\ \mathbf{D}(\mathbf{r'}) + \frac{1}{\varepsilon_0} \nabla' \cdot ( \kappa (\mathbf{r'}) \mathbf{D}(\mathbf{r'})  ) \nabla \Big \} \ g(\mathbf{r}, \mathbf{r'})\ d \mathbf{r}',
    \end{multline}
    where $ g(\mathbf{r}, \mathbf{r'}) = \frac{ e^{-j k_0 |\mathbf{r}-\mathbf{r'}|}}{ 4 \pi |\mathbf{r}-\mathbf{r'}|} $ is the scalar Green's function, $ \kappa(\mathbf{r'}) =\frac{\varepsilon(\mathbf{r'}) - \varepsilon_0 } { \varepsilon(\mathbf{r'}) }$ is the contrast ratio, $ \omega $ is the angular frequency, and $k_0$ denotes the free-space wave number.

    To solve Eq.~(\ref{vie}), we discretize the volume $ V $ (metal gates) with a tetrahedral mesh, expand $ \mathbf{D}(\mathbf{r}) $ using the Schaubert-Wilton-Glisson (SWG) basis~\cite{1984_SWG}, and do a standard Galerkin testing \cite{1984_SWG, Wilton1984Singularity}. SWG basis is a vector basis used to expand unknown electric flux density ($\mathbf{D}$) in each tetrahedron element (used to discretize the structure)~\cite{1984_SWG}. It ensures the normal continuity of the $\mathbf{D}$-field across the triangle face in the tetrahedron mesh naturally. %SWG basis is a linear function of space variables in each tetrahedron element. An SWG basis is associated with each triangle face in a tetrahedron mesh, and is defined in the space of two tetrahedrons attached to the triangle face.  It ensures the normal continuity of the D-field across the triangle face. 
    If the number of unknowns $ N $ is small, we can solve it using a full matrix. Otherwise, we could employ a fast solver to compress the dense matrix and solve it in $ O(N \log N + N N_{\text{iter}}) $ with an iterative solver or $ O(N \log N)$ time with a direct inverse \cite{YifanWang_TAP2022, 2018_Maiomiao_Direct}. After solving Eq.~(\ref{vie}), the scattered magnetic field can be obtained from the scattered electric field $\mathbf{E}^{sc}$.
	
	\subsection{Accuracy of VIE simulations}
	\begin{figure}[h]
	    \centering
	    \includegraphics[width= 3.375 in]{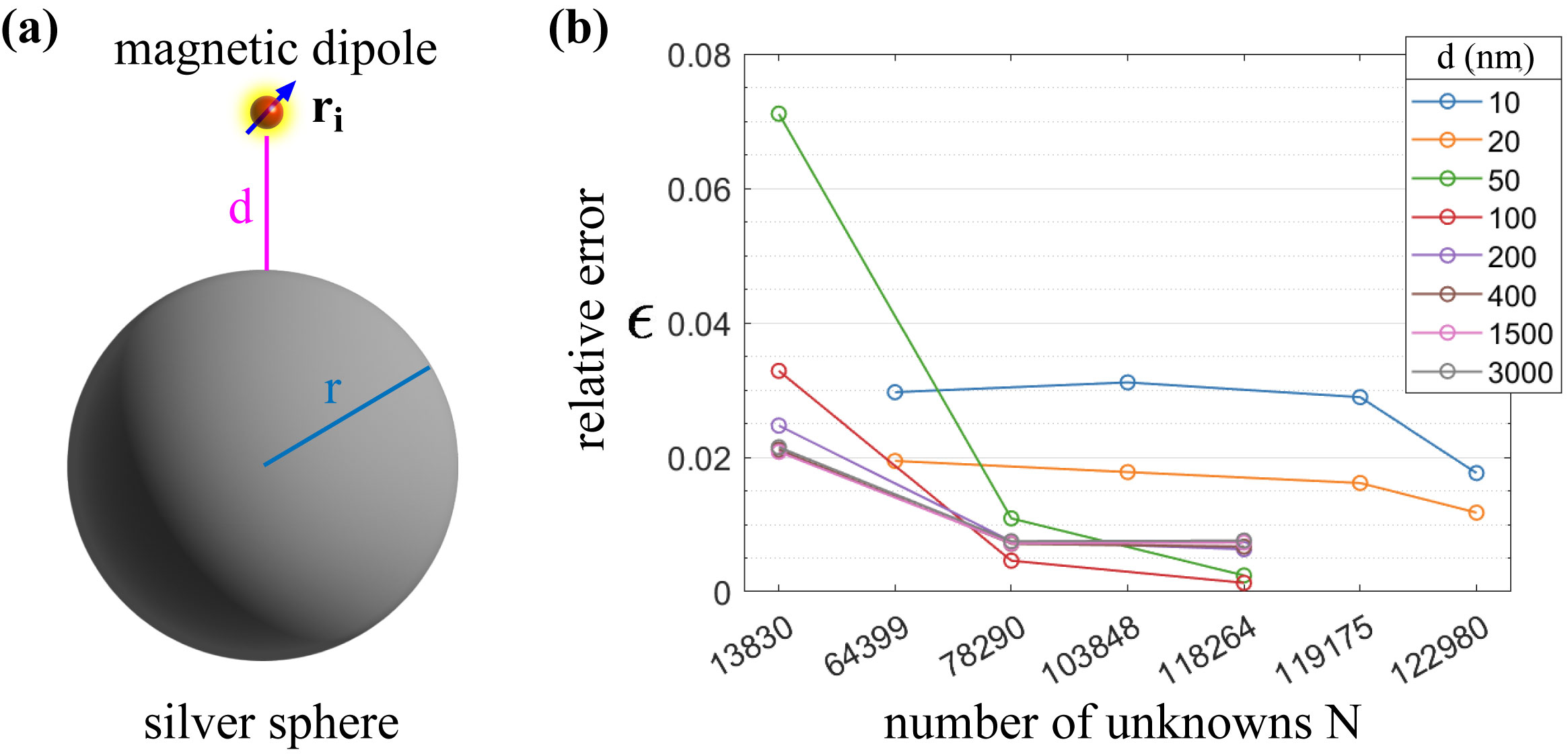}
	    \caption{Accuracy check of the VIEs simulations. (a) Schematic of the simulation. Magnetic dyadic Green's function at $\mathbf{r}_{i}$ is calculated from VIE simulations and analytical expressions separately. Here, we consider $r=500 \, \mathrm{nm}, \mathrm{Im} \, \varepsilon=10^9, \, \omega=2\pi \times 2.5 \, \mathrm{GHz}$ in this accuracy check. (b) Relative error $\epsilon$ of VIE simulated magnetic dyadic Green's functions with different $d$ and $N$.}
	    \label{fig:apfig2}
	\end{figure}
	
	\begin{figure}[h]
	    \centering
	    \includegraphics[width= 2 in]{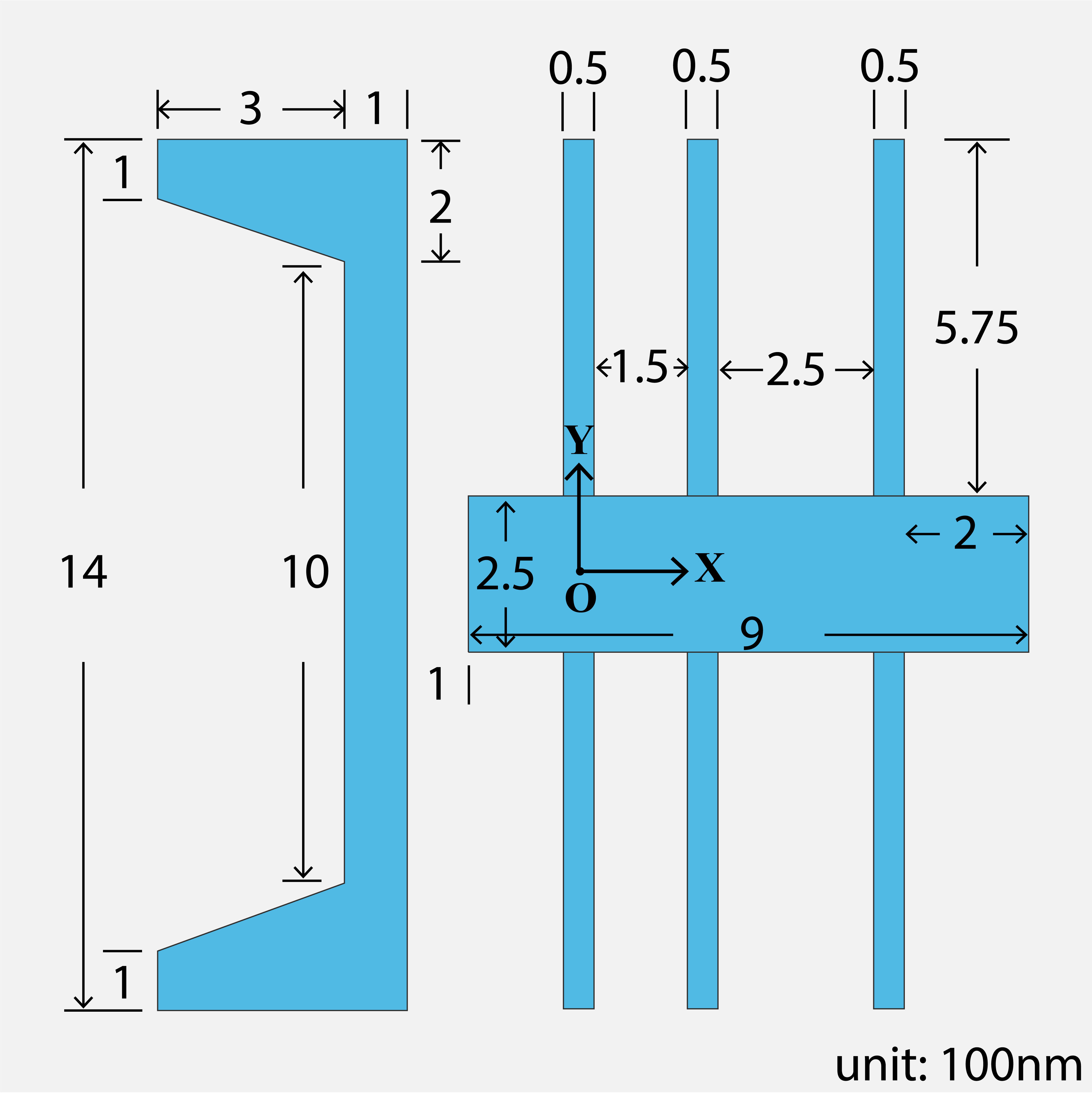}
	    \caption{Geometry of metal gates in the model device \cite{Huang2019fbenchmark}. The lengths are in 100nm unit. O is the origin of the Cartesian coordinates. Two qubits are considered to be at $(0,-25\,\mathrm{nm})$ and $(0,25\,\mathrm{nm})$ and at the same distance $d$ below the metal gates.}
	    \label{fig:apfig3}
    \end{figure}
	Here, we examine the accuracy of magnetic dyadic Green's functions obtained via VIE simulations. We compare the analytical \cite{tai1994dyadic} and VIE simulated Green's function $\overleftrightarrow{G}_m (\mathbf{r}_{i},\mathbf{r}_{i},\omega)$ in the vicinity of a silver sphere (Fig. \ref{fig:apfig2}(a)). Relative error $\epsilon$ of VIE simulated Green's function can be reduced to less than $1\%$ with an increasing number of meshes (Fig. \ref{fig:apfig2}(b)). In section \ref{section3}, we simulate $\overleftrightarrow{G}_m (\mathbf{r}_{i},\mathbf{r}_{i},\omega)$ and $\overleftrightarrow{G}_m (\mathbf{r}_{i},\mathbf{r}_{j},\omega)$ in the vicinity of metal gates in a quantum computing device (Fig.~\ref{fig:apfig3}), we use a refined mesh, and the relative error is estimated to be less than $5\%$.

	\section{Numerical Simulations of System Dynamics in the Liouville Space}\label{dnumerical}
	
	In this section, we discuss simulation methods for studying the dynamics of the two-spin-qubit system in the Liouville space \cite{yang2019engineering,Gyamfi_2020}. The Liouville representation is effective for solving the Lindblad master equation. In the Liouville space, a density matrix of size $N \times N$ is represented by a column vector of size $N^2 \times 1$. As a result, equations of motion for density matrices in the Liouville space can be solved by similar techniques that have been developed to solve equations of motion for state vectors in the Hilbert space. In the following, we present the density matrices and operators in the Liouville space used in our simulations. $\otimes$ represents the Kronecker product.
	
	In the Liouville space, we transform the two-spin-qubit density matrix $\rho_q$ (4 by 4 matrix) to a vector $| \rho_q \rangle$ (16 elements):
	
	\begin{equation}
	    | \rho_q \rangle = |\rho_{11}, \cdots, \rho_{14}, \cdots, \rho_{41}, \cdots, \rho_{44} \rangle ^T,
	\end{equation}
	where $\rho_{ij}$ is the element of the two-spin-qubit density matrix $\rho_q$.
	
	The commutator between Hamiltonian $\hat{H}$ and the two-spin-qubit density matrix $\rho_q$ is transformed to the super-operator acting on $| \rho_q \rangle$:
	
	\begin{equation}\label{neumannliouvillespace}
	    -i[\hat{H},\rho_q]= -i~(\hat{H} \otimes \hat{I} - \hat{I} \otimes H^T) ~ | \rho_q \rangle,
	\end{equation}
	where $\hat{I}$ is the identity matrix of the same size as Hamiltonian $\hat{H}$.
	
	The Lindblad super-operator in the Liouville space can be obtained based on the following transformations \cite{yang2019engineering}:
	
	\begin{equation}
	    \hat{\sigma}_{i}\rho_q(t)\hat{\sigma}_{j}^{\dagger} = \hat{\sigma}_{i} \otimes (\hat{\sigma}_{j}^{\dagger})^T |\rho_q\rangle,
	\end{equation}
	
	\begin{equation}
	    \rho_q(t) \hat{\sigma}_{i}^{\dagger}\hat{\sigma}_{j} = \hat{I} \otimes (\hat{\sigma}_{i}^{\dagger}\hat{\sigma}_{j})^T  | \rho_q \rangle,
	\end{equation}
	
	\begin{equation}
	    \hat{\sigma}_{i}^{\dagger}\hat{\sigma}_{j} \rho_q(t) = (\hat{\sigma}_{i}^{\dagger}\hat{\sigma}_{j}) \otimes \hat{I} | \rho_q \rangle.
	\end{equation}
    
    As a result, the Lindblad master equation (\ref{mainlindbladT}) with respect to $| \rho_q \rangle$ becomes:
    
    \begin{equation}
        \frac{d \, |\rho_q (t)\rangle}{dt} = \hat{\hat{L}}_{tot}(t) \ |\rho_q (t)\rangle,
    \end{equation}
    where $\hat{\hat{L}}_{tot}(t)$ contains both the unitary evolution component (Eq.~(\ref{neumannliouvillespace})) and the non-unitary components $\hat{\hat{L}}_r$, $\hat{\hat{L}}_{\phi}$. 
    
    The dynamics of $|\rho_q (t)\rangle$ can be calculated by splitting $[0,t]$ into $N$ intervals $\Delta t_1, \cdots , \Delta t_N$:
    
    \begin{equation}
        |\rho_q (t)\rangle = e^{\hat{\hat{L}}_{tot}(t_N) \Delta t_N} \cdots e^{\hat{\hat{L}}_{tot}(t_1) \Delta t_1} |\rho_q (0)\rangle.
    \end{equation}
    
	\section{Optimization Constraints}\label{optimizationconstraints}
	
	In Sec.~\ref{section5}, we apply Lindbladian engineering to diamond NV center system and silicon quantum dot system. The corresponding pulse optimization is a nonlinear optimization problem with constraints. Optimization constraints should be determined by actual experimental limits. In Sec.~\ref{subsection1}, we consider lower bound $\mathcal{X}_{l}=[0.1,0.1,0.1,0.1,0.1,0.1,0.1,-\pi,-\pi,-\pi]$ and upper bound $\mathcal{X}_{u}=[5,5,5,5,4,4,4,\pi,\pi,\pi]$ for pulse optimization in the NV center system. The initial point is $\mathcal{X}_i= [3.78,2.11,2.15,0.63,1.88,3.96,1.90,0,\pi/5,\pi/2]$ from experiments in reference \cite{PhysRevLett.124.220501}. In Sec.~\ref{subsection2}, we consider lower bound $\mathcal{X}_{l}=[0.1\pi,0]$ and upper bound $\mathcal{X}_{u}=[\pi,10]$ for pulse optimization in the silicon quantum dot system. The initial point is $\mathcal{X}_i=[0.8\pi,1.25]$.
	
	\section{Error Correction and Gate Fidelity}\label{errorcorection}
	
\begin{figure}[ht]
	    \centering
	    \includegraphics[width= 2.8 in]{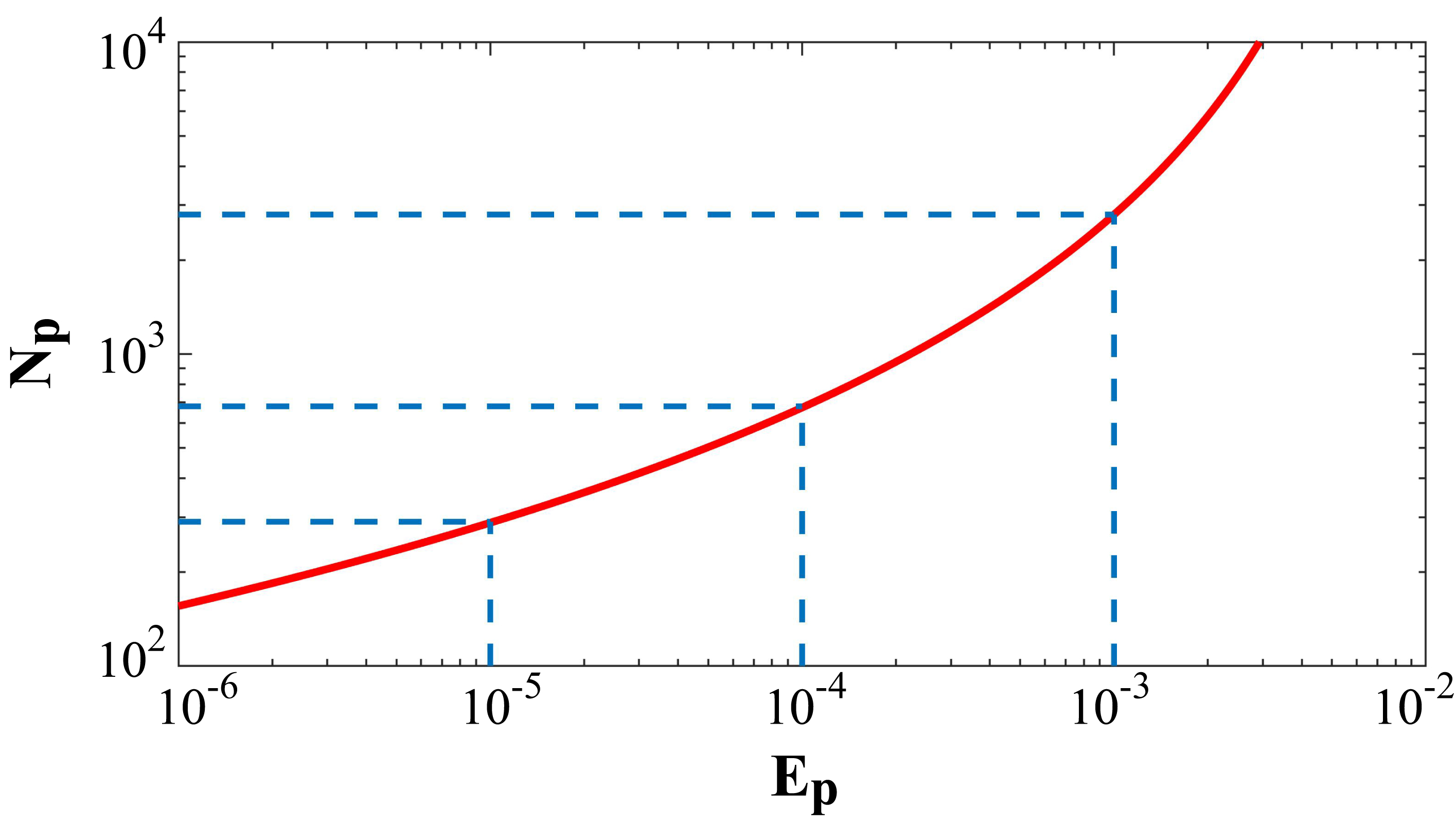}
	    \caption{Relation between physical qubit error rate $E_{p}$ and number of physical qubits $N_{p}$ for one logical qubit in Shor's algorithm implementation. We consider a logical qubit error rate $E_l = 10^{-15}$ and a surface code threshold $E_{T}=10^{-2}$ in this plot~\cite{PhysRevA.86.032324}.}
	    \label{fig:apfig1}
	\end{figure}

	In this appendix, we discuss the importance of physical quantum gate fidelity to quantum computer size. Logical qubits are the computational qubits for quantum algorithm realization. In surface code, one single logical qubit is constructed by multiple entangled physical qubits. The number of physical qubits $N_{p}$ needed for a logical qubit is sensitive to the error rate in physical qubits $E_{p}$, which is closely related to the fidelity of physical quantum gates $F$~\cite{PhysRevA.86.032324}. Figure~\ref{fig:apfig1} shows the relation between $N_{p}$ and $E_{p}$ in Shor's algorithm implementation based on the empirical formula~\cite{PhysRevA.86.032324}, where the logical qubit error rate is $E_l = 10^{-15}$, and the surface code threshold is $N_{p}=10^{-2}$~\cite{PhysRevA.86.032324}. We can find that reducing gate infidelity $\Delta F$ from $10^{-3}$ to $10^{-5}$ can lead to an $N_{p}$ decrease by a factor of $10$.
	
    \section{Analysis of the Average Gate Infidelity Induced by EWJN}\label{avgfidelity}
    In this appendix, we analyze the average gate infidelity $\overline{\Delta F}$ induced by near-field thermal and vacuum fluctuations. The main difference between $\overline{\Delta F}$ and $\Delta F$ is that all possible input states are considered in the average for $\overline{\Delta F}$, while only four computational basis states are considered in the average for $\Delta F$. $\overline{\Delta F}$ can be evaluated similarly as $\Delta F$ by considering the following average gate fidelity $\overline{F}$~\cite{PhysRevA.60.1888,nielsen2002simple,bowdrey2002fidelity,PhysRevB.103.L161409} instead of $F$ defined in Eq.~(\ref{equation5}):
    \begin{align}\label{avgf}
    \begin{aligned}
        \overline{F}&=\int d\psi \, \langle \psi | \mathcal{U}^\dagger_{\mathrm{CNOT}} \, \mathcal{E} (|\psi\rangle \langle \psi |) \, \mathcal{U}_{\mathrm{CNOT}} |\psi\rangle\\
        &=\int d\psi \, \mathrm{Tr}\, \left[ \mathcal{U}_{\mathrm{CNOT}}  |\psi\rangle \langle \psi | \, \mathcal{U}^\dagger_{\mathrm{CNOT}} \,\, \mathcal{E} (|\psi\rangle \langle \psi |) \right],
    \end{aligned}
    \end{align}
    where the integral considers all normalized input states $\psi$, $U_{\mathrm{CNOT}}$ represents the ideal CNOT gate, $\mathcal{E} (|\psi\rangle \langle \psi |) = \rho_q(|\psi\rangle \langle\psi|,t_f) $ represents the output density matrix of the noisy CNOT gate $\mathcal{E}$ when the input density matrix is $|\psi\rangle \langle \psi |$. Compared with $F$ in Eq.~(\ref{equation5}) where four computational basis states are considered in the average, $\overline{F}$ considers all input states in the average. The average quantum gate infidelity $\overline{\Delta F}$ induced by EWJN can be defined as the differences between the actual average CNOT gate fidelity $\overline{F}$ and the ideal counterpart $\overline{F_0}$ where EWJN is ignored, $\overline{\Delta F} = \overline{F_0} -\overline{F}$.

    \begin{figure}
        \centering
        \includegraphics[width=2.7in]{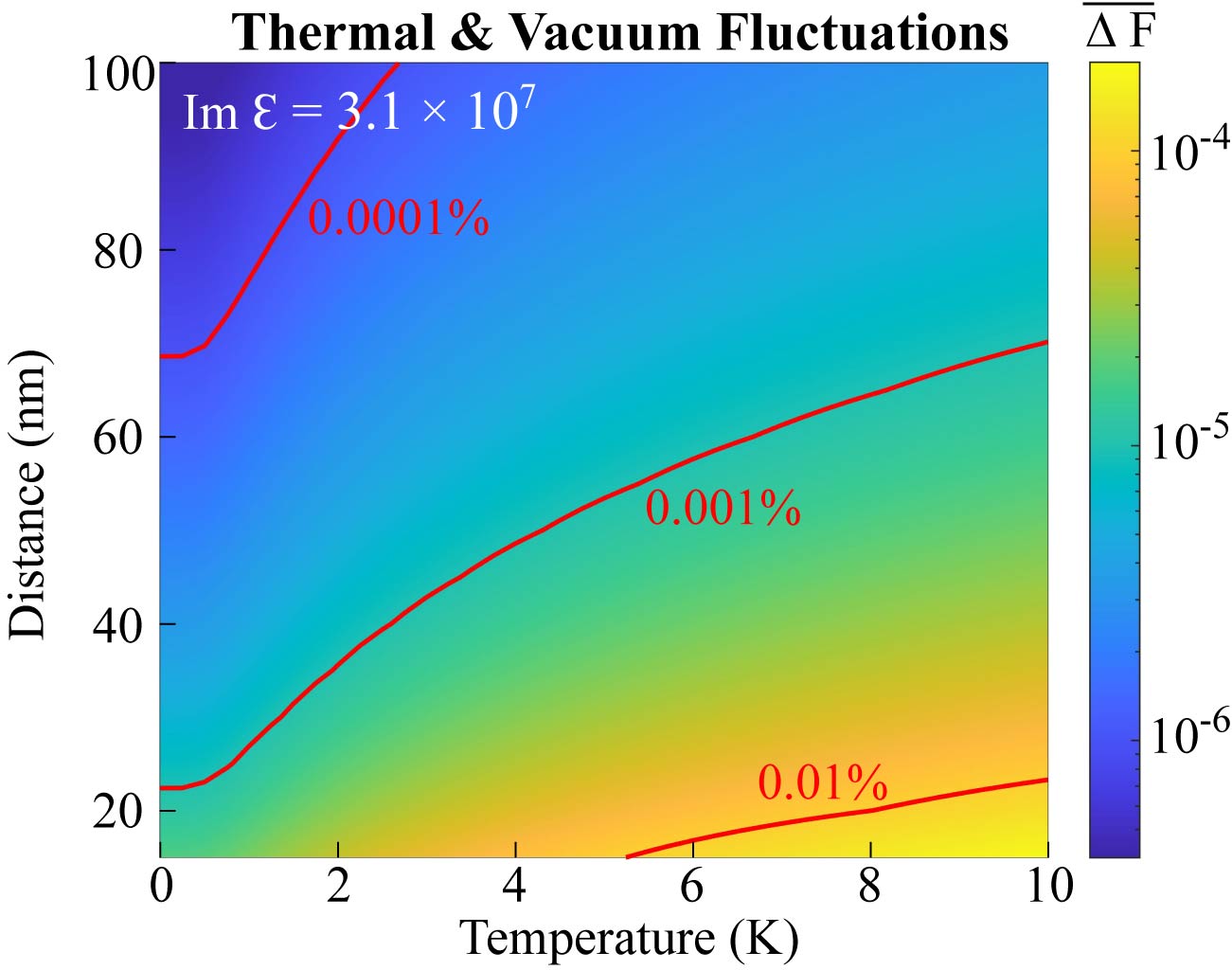}
        \caption{Average gate infidelity $\overline{\Delta F}$ induced by near-field vacuum and thermal fluctuations in a realistic quantum computing device based on the silicon double quantum dot (DQD) system. Dependence of $\overline{\Delta F}$ on temperature $T$ and distance $d$ at fixed $\mathrm{Im}\, \varepsilon$ is shown. Red lines represent constant contour lines of $\overline{\Delta F}$. Same parameters as in Fig.~\ref{fig:fig3}(d) are employed here.}
        \label{fig:apfigavgf_qd}
    \end{figure}
    
    Here, we use the following equivalent formula to evaluate Eq.~(\ref{avgf})~\cite{nielsen2002simple,PhysRevApplied.16.064031}:
    \begin{equation}
        \overline{F}=\frac{\sum_{i=1}^{\mathcal{D}^2} \mathrm{Tr}[\, \mathcal{U}_{\mathrm{CNOT}} \, \mathcal{U}_i^\dagger \, \mathcal{U}_{\mathrm{CNOT}}^\dagger \, \mathcal{E}(\mathcal{U}_i) \, ] +\mathcal{D}^2}{\mathcal{D}^2 \, (\mathcal{D}+1)},
    \end{equation}
    where $\mathcal{D}$ is the dimension of the quantum system ($\mathcal{D}=4$ for the two-qubit system), $\{\mathcal{U}_i\}_{i=1}^{\mathcal{D}^2}$ form the orthogonal basis of $\mathcal{D} \times \mathcal{D}$ unitary operators. For a two-qubit system, the 16 $\mathcal{U}_i$s can be expressed by the tensor products of Pauli matrices:
    \begin{equation}
        \mathcal{U}_i = \sigma_j \otimes \sigma_k,
    \end{equation}
    where $j, k \in \{0,1,2,3\}$, $\sigma_0=\overleftrightarrow{I}$ is the identity matrix, $\sigma_1,\sigma_2,\sigma_3$ are the Pauli matrices. In Eq.~(\ref{avgf}), $\mathcal{E}(\mathcal{U}_i)$ can be obtained from the system dynamics governed by the Lindblad master equation (\ref{mainlindbladT}). The simulations of $\mathcal{E}(\mathcal{U}_i)$ are performed in the Liouville space as described in Appendix \ref{dnumerical}. 
    
    \begin{figure}
        \centering
        \includegraphics[width=3in]{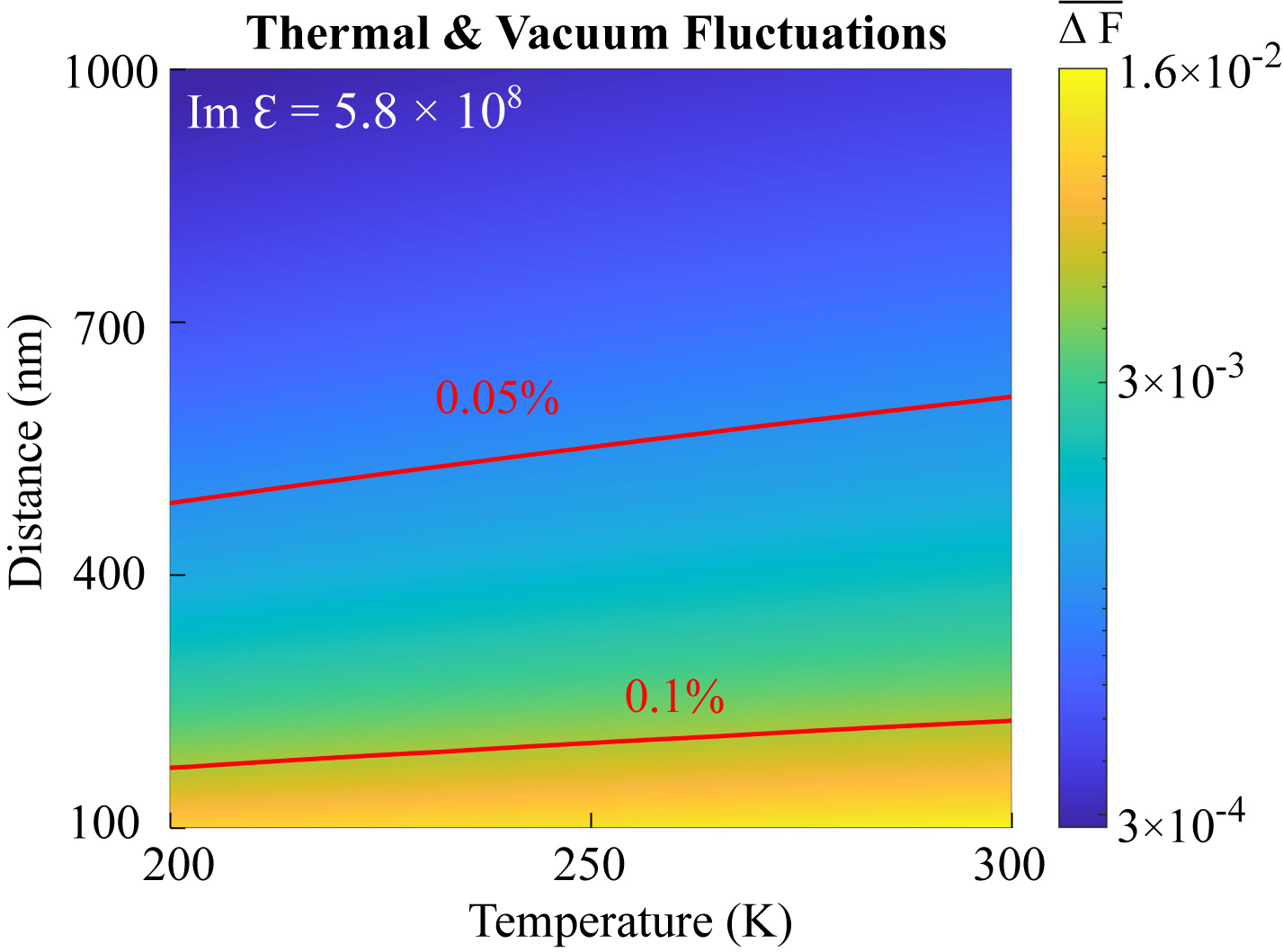}
        \caption{Average gate infidelity $\overline{\Delta F}$ induced by near-field vacuum and thermal fluctuations in a two-spin-qubit system based on the diamond NV center system. Dependence of $\overline{\Delta F}$ on temperature $T$ and distance $d$ at fixed $\mathrm{Im}\, \varepsilon$ is presented. Red lines represent constant contour lines of $\overline{\Delta F}$. Same parameters as in Fig.~\ref{fig:fig4}(c) are employed here.}
        \label{fig:apfigavgf_nv}
    \end{figure}
    
    In Fig.~\ref{fig:apfigavgf_qd}, we present the average gate infidelity $\overline{\Delta F}$ induced by near-field thermal and vacuum fluctuations in a two-qubit silicon DQD device (schematic in Fig.~\ref{fig:fig3}(a)). We employ the same parameters as in Fig.~\ref{fig:fig3}(d). The only difference %between Fig.~\ref{fig:apfigavgf} and Fig.~\ref{fig:fig3}(d) 
    is that we evaluate gate fidelity $F$ to obtain $\Delta F$ in Fig.~\ref{fig:fig3}(d), while we calculate average gate fidelity $\overline{F}$ to obtain $\overline{\Delta F}$ in Fig.~\ref{fig:apfigavgf_qd}. It is clearly shown that for EWJN in the silicon DQD system, the average gate infidelity $\overline{\Delta F}$ has close values and identical temperature and distance dependence as the $\Delta F$ evaluated in the main text. This result further facilitates considerations of EWJN in fault-tolerant quantum computing~\cite{PhysRevLett.122.080504,Sanders_2016} and comparison with randomized benchmarking~\cite{Emerson_2005,PhysRevLett.123.060501, PhysRevLett.109.080505,PhysRevA.89.062321,PhysRevA.77.012307,Wallman_2014}.
    
    In Fig.~\ref{fig:apfigavgf_nv}, we show the average gate infidelity $\overline{\Delta F}$ induced by near-field thermal and vacuum fluctuations in a two-qubit system based on the diamond NV center system (schematic in Fig.~\ref{fig:fig4}(a)). Here, we consider the same parameters as in Fig.~\ref{fig:fig4}(c). By comparing Fig.~\ref{fig:apfigavgf_nv} and Fig.~\ref{fig:fig4}(c), we find that for EWJN in the NV center system, $\overline{\Delta F}$ and $\Delta F$ are close and have identical dependence on temperature and distance from metallic control systems.
    
	\clearpage
	\bibliography{reference} % Produces the bibliography via BibTeX.

\end{document}